\pdfoutput=1

\documentclass[manuscript,acmsmall,screen]{acmart}
\AtBeginDocument{%
  \providecommand\BibTeX{{%
    \normalfont B\kern-0.5em{\scshape i\kern-0.25em b}\kern-0.8em\TeX}}}

\usepackage{graphicx} 
\usepackage{times}
\usepackage{latexsym}
\usepackage{multirow}
\usepackage{tabularx,tabulary}
\usepackage{amsmath}
\usepackage{breqn}
\usepackage{xcolor}
\usepackage{tikz}
\usepackage{comment}
\usepackage{adjustbox}
\usepackage{threeparttable}
\usepackage{times}
\usepackage{pdfpages}
\usepackage{subfig}
\usepackage{amsmath}
\usepackage{pgfplots}
\usepackage{hyperref} 
\usepackage{subfig}

\usepackage{enumitem}
\usepackage{longtable}
\usepackage{soul}
\usepackage{xcolor}

\newcommand{\changed}[1]{{\color{black} #1}}
\newcommand{\newchanged}[1]{{\color{black} #1}}

\begin{document}

\title{\changed{Online Self-Disclosure, Social Support, and User Engagement During the COVID-19 Pandemic}}

\author{Jooyoung Lee}
\email{jfl5838@psu.edu}
\affiliation{%
  \institution{Penn State University}
  \city{State College}
  \state{Pennsylvania}
  \country{USA}
}

\author{Sarah Rajtmajer}
\email{smr48@psu.edu}
\affiliation{%
  \institution{Penn State University}
  \city{State College}
  \state{Pennsylvania}
  \country{USA}
}

\author{Eesha Srivatsavaya}
\email{efs5377@psu.edu}
\affiliation{%
  \institution{Penn State University}
  \city{State College}
  \state{Pennsylvania}
  \country{USA}
}

\author{Shomir Wilson}
\email{shomir@psu.edu}
\affiliation{%
  \institution{Penn State University}
  \city{State College}
  \state{Pennsylvania}
  \country{USA}
}

\renewcommand{\shortauthors}{Lee, et al.}

\begin{abstract}
We investigate relationships between online self-disclosure and received \changed{social support and user engagement} during the COVID-19 crisis. We crawl a total of 2,399 posts and 29,851 associated comments from the \textbf{r/COVID19\_support} subreddit and manually extract fine-grained personal information categories and types of social support sought from each post. We develop a BERT-based ensemble classifier to automatically identify types of support offered in users' comments. We then analyze the effect of personal information sharing and posts' topical, lexical, and sentiment markers on the acquisition of support and five interaction measures (submission scores, the number of comments, the number of unique commenters, the length and sentiments of comments). Our findings show that: 1) users were more likely to share their age, education, and location information when seeking both informational and emotional support, as opposed to pursuing either one; 2) while personal information sharing was positively correlated with receiving informational support when requested, it did not correlate with emotional support; 3) as the degree of self-disclosure increased, information support seekers obtained higher submission scores and longer comments, whereas emotional support seekers' self-disclosure resulted in lower submission scores, fewer comments, and fewer unique commenters; 4) post characteristics affecting \changed{audience response} differed significantly based on types of support sought by post authors. These results provide empirical evidence for the varying effects of self-disclosure on acquiring desired support and user involvement online during the COVID-19 pandemic. Furthermore, this work can assist support seekers hoping to enhance and prioritize specific types of \changed{social support and user engagement}.
\end{abstract}




\maketitle

\section{Introduction}
A growing body of research sheds light on the adverse impacts of the COVID-19 pandemic on individuals' physical and mental well-being \citep{xiong2020impact, fancourt2021trajectories}. According to the CDC's Morbidity and Mortality Weekly Report \citep{czeisler2020mental}, U.S. adults from April to June of 2020 experienced a heightened prevalence of anxiety (25.5\% vs. 8.1\%) and depressive (24.3\% vs. 6.5\%) disorders when compared to the same period in 2019. \citet{reger2020suicide} suggests that these phenomena are attributable to several factors: financial concerns; social isolation; decreased access to community and religious support; and barriers to the acquisition of healthcare. In particular, a strong correlation between social isolation and degradation in psychological conditions has been studied \citep{douglas2009preparing, hossain2020mental}.

In the midst of the global epidemic, many have resorted to online support forums to cope with the impacts of COVID-19 while maintaining physical distance from others \cite{chew2020narrative}. Peer-to-peer interaction on the Web is known to reduce depression and enhance the quality of life \citep{wang2015eliciting, rains2009meta}. Particularly, it can facilitate increased social connectedness, feelings of group belonging \citep{naslund2016future} and guidance for important health care decisions \citep{ridout2018use}. To maximize these gains, support seekers tend to engage in self-disclosure, the voluntary sharing of personal information with others \citep{derlaga1987self}, more actively than non support seeking individuals \citep{wingate2020influence, lee2013lonely}. By providing more detailed descriptions of themselves and their situations, they can acquire the increased quality and quantity of advice received from peers \citep{pan2018you}. 



\changed{Although prior efforts have attempted to understand individuals' self-disclosure behaviors in online health communities (OHCs) \citep{valizadeh2021identifying, balani2015detecting}, most focused on OHCs for chronic diseases and mental health disorders. Little is known about the types of personal and private information that people disclose when they share experiences of global health crises online. Given the novelty and impact of the COVID-19 outbreak, it has been suggested that the response in terms of online self-disclosure has been unique \citep{nabity2022ll, zhen2021college}. Specifically, levels of voluntary sensitive information disclosure have intensified during the COVID-19 crisis \citep{blose2021study, umar2020study}. Recent research has investigated how particular populations such as pregnant women \cite{lei2022pregnant}, Chinese people \cite{dou2022empathy}, or college students \cite{zhen2021college} engaged in online self-disclosure to obtain social support during  COVID-19. This motivates us to explore how the pandemic has contributed to new practices for personal information sharing in a broader context.}

\newchanged{Recently, OHCs have begun to emerge within social network sites (SNSs), facilitating more interactive communication between OHC members. This enabled members to vividly express their attention and care through various SNS features such as commenting or pressing `like' buttons. While several studies \cite{yang2019channel, li2021predicting, de2014mental} have studied relationships between online self-disclosure of support seekers and corresponding responses (e.g., supportive vs. unsupportive responses) from other members, none of them differentiated the categories of support sought and provided by support seekers and givers, respectively. Identification of support types from the content is necessary because the optimal form of social support is one that aligns with the individual’s needs \cite{mattson2011health}. There exist two studies \cite{wang2015eliciting, huang2016examining} that distinguished these support types, but their findings were inconsistent due to variations in study populations, contexts, and platforms. 

User engagement attributes, such as the number of comments or likes received from friends, are additional factors that influence the social presence and satisfaction of support-seeking individuals from online self-disclosure \cite{derlega2008self,berg1987responsiveness, pinquart2000influences, cheikh2014like}. Particularly, high levels of audience engagement encourage them to share information more actively \cite{ernala2018characterizing, sultan2021user}. Still, the impact of self-disclosure on user engagement rates remains unknown. Understanding this dynamic is critical, as user engagement levels play a significant role in the social benefits individuals can derive from readers of their self-disclosing posts \cite{de2017gender}.

Research has revealed a connection between linguistic characteristics of self-disclosing content and the attainment of social support and user engagement. For example, the increased readability level and the number of words positively affect the quantity of social support the post obtains through comments \cite{chen2020linguistic}. Also, while posts that are easy to read and contain more than 31 words and many hashtags were likely to achieve higher performance of engagement and awareness \cite{gkikas2022text}, some topics or emotions in the post may cause users to be uninterested in responding \cite{he2020dynamics, gerbaudo2019positive}. In light of the changes in people's posting behaviors during COVID-19, there is a need for reevaluation of relationships between language cues and social support and user engagement acquisition.
}

 \changed{In the light of the above discussion, we aim to identify correlations between support seekers' public discourse and the receipt of \textbf{social support} and \textbf{user engagement} within the \textbf{r/COVID19\_support} subreddit\footnote{\url{https://www.reddit.com/r/COVID19_support/}}, as well as the role of linguistic queues of posts in these interactions. Specifically, this study is guided by three research questions:
\begin{enumerate}
    \item What types of personal information do \newchanged{COVID-19} support seekers disclose based on requested social support?
    \item How is online self-disclosure related to \newchanged{COVID-19} support seekers receiving requested social support and user engagement?
    \item What characteristics of self-disclosing posts are related to the requested social support and user engagement acquisition \newchanged{in the context of COVID-19}?
\end{enumerate}
}

\newchanged{
\noindent Of several categories of social support, we focus on emotional and informational support as they have received the most theoretical and empirical attention. For self-disclosure, we investigate 11 personal information types that are related to factual information about the author (e.g., age, education, employment status). 
We manually annotate all posts for types of social support and personal information. For replies to these posts, we develop deep learning models that identify social support provision. User engagement rates are quantified via five components for RQ2. We also examine five textual attributes, the most common stylometry- and content-wise features \cite{umar2019detection, wang2012stay}, of self-disclosing content for RQ3.}

\newchanged{Our findings show that `r/COVID19\_support' members tended to alter their personal information-sharing practices based on the support categories they ask for. For instance, their age, education, or location information disclosure became more frequent when seeking both informational and emotional support compared to pursuing either one. We also observe that online self-disclosure had a contradicting effect on the receipt of informational vs. emotional support. Specifically, self-disclosure increased the chances of users obtaining solicited informational support, but the same did not hold for emotional support. Lastly, we discover that textual attributes encouraging \changed{the provision of social support and user engagement} differed significantly depending on support types.
 }

 
The main contributions of our research include the following:
\begin{itemize}
 
 \item 
 Our analyses with 11 fine-grained informational self-disclosure categories shed light on what kinds of personal information was disseminated within the \textbf{r/COVID19\_support} subreddit and how users alter their information sharing practices based on their needs. \changed{The results can be useful in predicting support-seeking individuals' self-disclosing behaviors during future emergencies and health crises.}
 
 \item \newchanged{We discover an unequal distribution of both rewards and losses from self-
disclosure depending on solicited support categories during COVID-19. The current study
raises awareness of this vulnerability and further invokes the development of technological
interventions that could help all individuals make better information-sharing decisions.}
 
 \item 
 Our findings \newchanged{regarding language and content of online self-disclosure in the need for social support} contribute to prior research by providing a holistic picture of the factors associated with these response phenomena and offering insights for adequate response acquisition strategies. \changed{It further assists social media designers and moderators facilitate more supportive and constructive interactions in online spaces.}

\end{itemize} 

\section{Related Work}

\subsection{Online Self-Disclosure}
Self-disclosure is `the process of making the self known to other persons' \cite{jourard1958some, joinson2007self}, which requires a speaker to voluntarily expose data about themselves. In particular, the degree of self-disclosure is reported to have intensified in computer-mediated communication interactions when compared to face-to-face discussions \cite{schouten2009experimental,tidwell2002computer}. The observed phenomenon may be explainable by ample opportunities for self-expression and social connection as a result of the rapid development of social media platforms, bringing in a wealth of user-generated content \cite{aaltonen2016cumulative}. When their content contains information about authors, topics of self-disclosure can be grouped into two categories, objective and subjective self-disclosure. Objective self-disclosure is related to factual information about an author (e.g., age, marital status, or employment status) whereas subjective self-disclosure includes intrinsic states of an individual (interests, opinions, and feelings) \cite{umar2019detection, lee2021digital}.

While engaging in online activities, users evaluate the expected rewards and risks associated with their information disclosure \cite{diaz2020preventative}. An estimation of the prospective gratifications and risks of revealing private information to others is commonly named as the \emph{privacy calculus} by privacy scholars \cite{dienlin2016extended, li2010understanding}. Positive societal rewards that users anticipate include receiving emotional and psychological comfort \cite{utz2015function, bazarova2015social, bekalu2019association}, maintaining good relationships with existing ties \cite{alloway2014facebook, sawyer2012impact}, and establishing new connections \cite{krasnova2010online}. These benefits explain the intensified self-disclosure within support communities. Still, oversharing with both strangers and acquaintances is not recommended. Based on analyses of Facebook users, practices of public discourse with minimal regard may result in damaged public reputation, the leakage and use of personal data by third parties, or hacking and identity theft \cite{debatin2009facebook}. Moreover, users sometimes feel remorse for sharing sensitive topics or content with negative sentiments \cite{madden2012privacy, wang2011regretted}, as well as location sharing \cite{patil2012reasons}. 

Apart from pursuing perceived rewards, public discourse can also occur depending on others’ disclosures \cite{acquisti2012impact, rajtmajer2016constrained}. \citet{acquisti2012impact} found that participants who were informed that other participants had disclosed sensitive data were more inclined to share personal information than those who were told nothing. Anonymity also encourages people to participate in online self-disclosure \cite{ma2016anonymity, postmes2001social}. These factors are not inherently flawed, but they can mislead Internet users into underestimating the costs of oversharing and leading to regrettable actions.

\subsection{Self-Disclosure during the Pandemic} 
Not only people's daily routines but also their online activities have been altered by the Coronavirus pandemic. Prior research has revealed a 61\% increase in the usage of social media platforms during the pandemic \citep{nabity2020inside}. The increased traffic could be a result of the practices of social distancing and mandatory quarantines, as people reach to online resources such as video-conferencing and social media platforms to continue their social and professional activities virtually \citep{lowenthal2020thinking, cinelli2020covid}. Although Information and Communications Technology (ICT) has aided consumers in effortlessly adapting to online learning or remote working, recent studies have suggested that pandemic breakouts harm consumer behavior to a large extent \cite{miri2020panic, naeem2021social}. One survey conducted by Rajab and Soheib \cite{rajab2021privacy} revealed that approximately 91\% of study respondents felt uncomfortable using webcams in online classes due to privacy concerns (88.4\%) and anxiety issues (64.4\%). Excessive online activity is reported to cause `corona fatigue’ which is associated with several problems including poor social media sleep hygiene, compulsive use, mental or physical health deterioration \cite{king2020problematic, dhir2021dark}.

Prior to the global pandemic, users tended to actively post their physical experiences on social media encouraging social connectedness to others \cite{kim2021contribution}. However, with the COVID-19 epidemic, countries around the world imposed strict restrictions on outdoor activities, socializing, and gatherings. Hence, topics of social media content have naturally shifted as well. Users are more involved in sharing personal health information (PHI) during the pandemic \citep{saha2020social}. At the same time, they are prone to self-censor their posts by measuring societal impacts of the contents \citep{nabity2020inside}. For example, contents that people regularly disclosed before the outbreak - such as going out for food, throwing a party, traveling, spending time outside - have become unfavorable, as they seem to violate public health guidance. These findings have brought to light a transformation of perceptions of what constitutes sensitive or private information. According to \citet{blose2021study}, overall personal information sharing practices have significantly intensified during the COVID-19 crisis. So far, most research is mainly focused on general audiences in SNS contexts, whereas empirical research into members of a particular COVID-19 support group is lacking.

\subsection{Social Support in Online Communities}
Individuals' experiences of being cared for, responding to, and being supported by people in their social group are referred to as social support \cite{cobb1976social}. Humans are social animals, and hence pursue social interactions in order to satisfy their social needs for belonging and support \cite{maslow1981motivation}. A positive influence of social support on both health and psychological health is widely recognized in the scientific literature. One of the early findings \cite{blazer1982social} proved that impaired perceived social support can adversely affect the individuals' health. Likewise, young people with lower quality social support tend to lack psychological well-being, including depressive or suicidal symptoms \cite{rigby1999suicidal, hefner2009social,chao2011managing}.

With the emergence the Internet, users can build intimate ties with others and gain practical information or emotional comforts without their physical presence \cite{eastin2005alt}. According to \citet{coulson2005receiving}, cyber spaces provide greater opportunities for a user to open up more freely and discuss sensitive topics with less risk than conventional face-to-face communication. Furthermore, as the general public gains confidence in the use of computer-mediated communication technologies, online support groups are rapidly growing \cite{white2001receiving}. If a user receives useful assistance from friends on online support groups, he or she may feel compelled to reciprocate, which makes more information available online \cite{crocker2008creating}.
That said, social support has become part of primary social values that individuals can obtain from an online community \cite{liang2011drives, nambisan2011information}.

According to the Social Support Behavioral Code \citep{cutrona1992controllability}, there are five different types of social support: informational support (providing practical information or advice), instrumental support (expressing their willingness to provide tangible assistance), esteem support (boosting respect and confidence in abilities by acts), network support (developing belonging to a group of people with similar concerns or experiences), and emotional support (communicating love, encouragement, or empathy). Emotional support and informational support have been most thoroughly explored in the existing works \citep{jaidka2020report, wang2012stay, biyani2014identifying}.

\subsection{Self-Disclosure \changed{, Social Support and User Engagement}}

Social support can be successfully exchanged when a support provider can provide adequate support and when a support recipient is willing to self-disclose \cite{trepte2018mutual}. Discussion forums intended specifically for support are reported to contain a higher degree of self-disclosure than general discussion forums \cite{barak2007degree}, and prior work shows that support seekers' self-disclosure increases perceived social support \cite{wingate2020influence, andalibi2016understanding, lee2013lonely}. More precisely, support seekers' openness tends to increase the amount of support they receive because it encourages their listeners to understand their needs and provide appropriate assistance \cite{huang2016examining}. 

Prior literature has sought to distinguish different types of social support in an attempt to identify their individual correlations with public discourse. Among five social support categories \citep{cutrona1992controllability}, emotional support (communicating love, encouragement, or empathy) and informational support (providing practical information or advice) have been most thoroughly explored. \citet{huang2016examining} recruited 333 Facebook users and confirmed a positive effect of online self-disclosure on both types of support. Two studies \cite{wang2015eliciting, yang2019channel} investigated the dynamics of social support exchange in online cancer support groups by incorporating automated measurements of self-disclosure and provision of support. According to \citet{wang2015eliciting}, self-disclosing text containing writers' negative emotions and experiences increases emotional support, whereas neither of those positively influences the provision of informational support. They also suggest that these observations are partially affected by perceived social needs. However, \citet{wang2015eliciting} did not explore underlying linguistic and topical attributes of content and their impacts on the receipt of solicited assistance. Moreover, their model suffers from low RMSE. Here, we provide a robust analysis on the dynamics of social support exchange through the lens of self-disclosure.

As many people turn to the internet for information and connection with like-minded peers, a large volume of digital data is continuously generated and leveraged by scholars to study social phenomena \cite{ahmed2020covid, bender2011seeking, de2012content} or its relationship with social support provision \cite{rains2016language, warner2018acquisition} and user engagement (i.e. number of likes, comments) \cite{cunha2016effect, aldous2019view}. Understanding what inspires people to respond to postings provides insight for how to keep them involved in their communities. A recent study \cite{chen2020linguistic} investigated the impact of different linguistic markers of the posts on social support exchange and discovered that posts' readability and spelling promote support provision. \changed{Also, \citet{de2017language} and \citet{sharma2018mental} attempted to identify whether linguistic accommodation affects social support acquisition across a variety of mental health related subreddits.}
User engagement patterns may also vary depending on the posting format and content. For example, according to \citet{molina2020content}, emotional appeal in Facebook posts achieves higher interaction in comparison to informational content. Similarly, the usage of hashtags and images have been shown to be useful in capturing user attention and increasing content engagement \cite{salomon2013moving,cvijikj2011effect}. Although prior work (e.g., \citet{pan2017you, wang2014social}) has explored how post content affects viewers' engagement rates specifically in online support forums, little is known about the role of self-disclosure, a core element of support-seeking content, on the receipt of \changed{desired social support and user engagement}. \citet{pan2018you} examined message length, emotion words, and cognitive processing words as linguistic criteria for interaction participation in support-giving messages and found that self-disclosure was linked to greater degrees of interaction involvement. However, unlike our work, they did not take into account social support receipt. \changed{\citet{andalibi2018social} analyzed audience's responses to sensitive disclosures of those who experienced sexual abuse, but failed to cover diverse personal information types while focusing on the role of anonymity.} 

\section{Manual Annotation of Self-Disclosure and Support-Seeking for Posts}\label{sec:manual}
In this section, we describe our data collection, self-disclosure annotation, and social support identification processes for the 2,399 posts in our dataset. Automated labeling of 29,851 comments that respond to these posts is explained in Section \ref{sec:automated}. 

\begin{table*}[]
{\small
\begin{tabular}{|c|c|}
\hline
\textbf{Statistics}                 & \textbf{N}       \\ \hline
Total number of posts           & 2,399  \\ \hline
Total number of comments        & 29,851 \\ \hline
Average posts per user          & 1.62   \\ \hline
Mean length of posts (words)    & 248.23 \\ \hline
Average scores per post        & 25.91  \\ \hline
Average comments per post       & 12.82  \\ \hline
Average comments per user       & 4.14   \\ \hline
Mean length of comments (words) & 70.45  \\ \hline
Average scores per comment     & 4.51   \\ \hline
\end{tabular}
}
\caption{Descriptive statistics of crawled dataset} 
\label{tab:descriptive}
\end{table*}

\subsection{Data Collection}
Among several social media platforms such as Facebook, Instagram, Reddit, or Youtube, we selected Reddit as a primary resource \newchanged{for several reasons. First, it is one of the largest aggregates of user-generated content where a variety of subgroups for any and every interest are supported \citep{anderson2015ask}. Each subreddit is monitored by moderators who filter out irrelevant content, enabling researchers to identify relevant groups and retrieve posts with minimal noise. Moreover, unlike most social networks which require users to maintain their offline identities, Reddit supports pseudonymity where users can create multiple temporary identities via throwaway accounts. The ability to remain anonymous may lower individuals’ perceived vulnerability to stigmatized content, facilitating more honest disclosure. Lastly, Reddit offers its own API, simplifying the process of collecting data.} 

\changed{Among few subreddits designed for \newchanged{COVID-19 specific social support exchange}, such as r/Coronavirus or r/COVID19}, we target the r/COVID19\_support subreddit \newchanged{because it was the largest community at the time of investigation, actively facilitating both informational and emotional support exchange.} The r/COVID19\_support subreddit was created on February 12th, 2020 and now has 32.5K users in total. We collected all public posts uploaded between February 12th, 2020, and February 16th, 2021 using PRAW, the Python Reddit API Wrapper.\footnote{https://praw.readthedocs.io/en/latest/} Due to the limitations of the official Reddit API, a maximum of 1,000 posts were downloaded per day. Since our primary interest is support-related content, we only crawled posts under the `Support' flair.\footnote{A flair is a tag that may be applied to threads, allowing users to identify the category to which the posts belong and assisting readers in filtering particular types of messages depending on their preferences.} The accompanying comments to these posts were retrieved subsequently. 
The resulting total number of posts and comments that were not deleted or removed were 2,399 and 29,851, respectively. Additional descriptive statistics of our crawled dataset can be found in Table \ref{tab:descriptive}.

\begin{table*}[h!]
\renewcommand{\arraystretch}{1.3}
{\small
\begin{adjustbox}{width=\textwidth,totalheight=\textheight,keepaspectratio}
\begin{tabularx}{\textwidth}{|l|c|X|}
\hline
\textbf{Categories}                           & \textbf{Sub-categories}     & \textbf{Description}                                                                                                      \\ \hline
\multirow{7}{*}{Demographic}         & Age                & Sharing a post that directly implies information about one’s own age                                                                          \\ \cline{2-3} 
                                     & Race/ethnicity     & Sharing a post that directly implies information about one’s own race or ethnicity such as being Black, White, Hispanic, etc.                 \\ \cline{2-3} 
                                     & Gender             & Sharing a post that directly implies information about one’s own gender                                                                       \\ \cline{2-3} 
                                     & Marital status     & Sharing a post that directly implies information about one’s own marital status such as being single, married, separated, divorced or widowed \\ \cline{2-3} 
                                     & Education          & Sharing a post that directly implies information about one’s own education level                                                              \\ \cline{2-3} 
                                     & Employment status  & Sharing a post that directly implies information about one’s current employment status                                                         \\ \hline
Location                                     & Location           & Sharing a post that directly implies information about one's own location such as postal code, street address, city, or state                              \\ \hline
Health information                   & Health information & Sharing a post that directly implies information about one's own physical or mental health condition, diagnosis, or prescriptions                  \\ \hline
\multirow{3}{*}{Personal identifier} & Phone number       & Sharing a post that directly implies information about one’s own phone number                                                                 \\ \cline{2-3} 
                                     & Email address      & Sharing a post that directly implies information about one's own email address                                                                \\ \cline{2-3} 
                                     & SSN                & Sharing a post that directly implies information about one's own social security number                                                       \\ \hline
None                                 & No disclosure      & Sharing no information about the above 11 personal information types                                             \\ \hline
\end{tabularx}
\end{adjustbox}
}
\caption{Descriptions of personal information categories} 
\label{tab:intruction}
\end{table*}

\subsection{Annotation for Self-Disclosure}
A body of work has emerged around automated detection of self-disclosed personal information in text through advanced natural language processing (NLP) \citep{squicciarini2020tipping, umar2019detection, bak2014self}. State-of-the-art approaches present either binary classification to capture the presence/absence of self-disclosure in general \citep{umar2020study}, or label broad categories of disclosure (e.g., informational vs. emotional) \citep{akiti2020semantics, dadu2020bert}, but do not comprehensively afford fine-grained categorical labels. Hence, we manually annotated our 2,399 collected posts according to 11 objective personal information categories derived from existing work \citep{lee2021digital}. We also added a `Health Information' category, which is known to have surged due to the pandemic \citep{saha2020social}. Table \ref{tab:intruction} provides a description of these 11 categories. Each post was categorized as representing one or more types of personal information or was labeled `No Disclosure' if none of these types were present.

Complete labeling for self-disclosure was performed by two authors of the paper. A detailed annotation instruction with descriptions and examples of each category was distributed. To ensure the reliability and consistency of the labeling, the two annotators initially annotated the first 50 instances of the corpus and calculated Cohen’s Kappa values. Results of calculated Kappa scores for each category are as follows: Age (0.751), Race/Ethnicity (1), Gender (0.539), Marital Status (0.734), Education (0.729), Employment Status (0.851), Location (0.558), Health Information (0.694), \changed{Phone Number (N/A), Email Address (N/A), SSN (N/A),} and No Disclosure (0.593). Three categories such as `Phone Number', `Email Address', and `SSN' are \changed{marked as N/A} because there were no labeled posts in these categories. Overall scores except for `Gender', `Location' and `No Disclosure' were above 0.6, indicating a good agreement rate. We assume lower values for some categories including `Gender' and `Location' may be explained by a lack of inclusive examples in the provided guideline. For instance, online users tend to disclose their age and gender together like `32F', but this expression was excluded from our annotation scheme. Similarly, while some people may label `I am from LA' as a location sharing post, others may reckon that it is too vague to categorize it as location because it is uncertain if a post author still lives in LA or not.

\subsection{Annotation for Support-Seeking Posts}
\begin{table*}[]
{\small
\centering
\begin{tabularx}{\textwidth}{|l|X|X|}
\hline
\textbf{Categories}            & \textbf{Description}                                                                                                                              & \textbf{Examples}                                                                                                                                                                                                                                                                                                                                                                      \\ \hline
Emotional support     & The post is seeking sympathy, caring, or encouragement.                                                                                  & \begin{itemize}\item This has been really tough for me. Does anyone have words of encouragement? \item I needed to vent out to feel better. Thanks for reading. \item Is anyone dealing with similar feelings living at home during this? I’ve just been feeling kind of down lately since most of my friends have been living on their own during Covid. \end{itemize}\\ \hline
Informational support & The post is seeking specific information, practical advice, or suggesting a course of action. & \begin{itemize}\item Do you know of a resource that can help children during deployment? \item Is it safe to go to school? We don't have an online option. \item What else protection do I need?  \end{itemize}  \\ \hline
\end{tabularx}
}
\caption{Descriptions of support-seeking behaviors} 
\label{tab:support_definition}
\end{table*}

The curation of robust labeled data for training AI-driven models to recognize support-seeking text in user-generated content is a persistent challenge. Recently, \citet{wang2021cass} implemented convolutional neural network (CNN)-based text classification with 2,300 labeled posts to distinguish informational support-seeking posts and non-informational seeking posts. Although their model achieved robust performance (accuracy: 0.86 / F1: 0.87), they did not release their dataset publicly. Hence, we manually annotated 2,399 posts for two types of social support (informational support vs. emotional support). These manual labels serve as the most reliable for the purposes of our study and, we suggest, may serve as important training data in future work developing automated approaches for this task.

In order to maintain the consistency between manual labels (for posts) and automated labels (for comments; described in Section \ref{sec:automated}), annotation instructions were based on Jaidka's work \cite{jaidka2020report}.\footnote{ Their labeling instructions can be found in this Github repository: \url{https://github.com/kj2013/claff-offmychest}.} The `General support' category was excluded from our instructions. We operationalize support as displayed in Table \ref{tab:support_definition}.


Given the provided instructions, two authors of the paper completed the labeling task. They were asked to label the first 100 instances for a Kappa value measurement. Cohen's Kappa scores for each category were used as a measurement. Overall, we observed good agreement among labelers: support (0.823), informational support (0.76), and emotional support (0.732). Upon confirmation of high agreement rates, the two annotators completed the annotation of the remaining 2,299 posts.

\begin{table}[]
\centering
{\small
\begin{tabular}{|c|c|}
\hline
\textbf{Label}                 & \textbf{N (\%)}       \\ \hline
Support               & 3226 (25.28) \\ \hline
Informational support & 1240 (9.72)  \\ \hline
Emotional support     & 1006 (7.88)  \\ \hline
\end{tabular}
}
\caption{Statistics of AAAI data} 
\label{tab:supportdata}
\end{table}

\section{Automatic Classification for Support-Giving Comments}
\label{sec:automated}
Following, we detail our approach to detect types of support offered in comments in our dataset. For model training, we utilize the publicly available dataset \citep{jaidka2020report}. Since the distribution of the dataset is highly imbalanced (particularly for informational and emotional support), we separated the support-giving comment classification task into two subtasks: 1) classify if a comment is providing any support (\emph{supportiveness classification}); 2) of the comments that give support, classify if a comment is offering informational support or emotional support (\emph{support type classification}).

\subsection{Supplementary Data for Model Training}

The Reddit dataset \cite{jaidka2020report} contains 12,860 labeled comments and 5,000 unlabeled comments that are collected from two subreddits, `r/CasualConversation' and  `r/OffMyChest'. The former subreddit is a sub-community where users are encouraged to casually share what is on their minds, and the latter is a support-based community where users can disclose deeply emotional matters. The topics of the corpus are restricted to relationships, with the following tags: `wife'; `girlfriend'; `gf'; `husband'; `boyfriend', and `bf'. 

The original dataset consists of six gold standard labels for each sentence: emotional disclosure; information disclosure; support; general support; information support; and emotional support. For the purpose of this paper, we only use labels of support, emotional support, and information support.

Supposedly, all sentences that are labeled as either `informational support' or `emotional support' should be labeled as `support'. However, we discovered 10 sentences that are labeled as `informational support', but not as `support'. We considered these examples as outliers and excluded them. The statistics of the data are detailed in Table \ref{tab:supportdata}.

\subsection{Task 1. Supportiveness Classification}

\textbf{Method} This task uses the `support' gold label to filter out comments that do not provide any support. Although the `support' label is not as skewed as the other labels, it is still imbalanced (ratio 3:1). In our case, `no support' is the majority class (n=9,634), and the positive class, `support', is the minority class (n=3,226). Models that are trained with unbalanced datasets are prone to predict tricky instances as the majority class because their goals are to minimize error rates. There are multiple techniques to mitigate the effect of class imbalance, such as oversampling the minority class or undersampling the majority class. Oversampling techniques balance the class distribution by randomly duplicating examples in the minority class, whereas undersampling involves deleting samples from the majority class. Cost-sensitive learning can also be used by assigning a higher cost for misclassification of the minority class \citep{elkan2001foundations}. This method is implemented by multiplying the loss of each example by a certain variable. Unlike sampling approaches where the data distribution has to be directly modified, a cost-sensitive learning technique is achieved during the process of model training. \citet{madabushi2020cost} has highlighted the effectiveness of incorporating cost-sensitivity into BERT when it comes to enhancing model generalization. As an experiment, we train our proposed model using two approaches: one that applies cost-sensitive learning and the other that does not employ any special techniques. We implement cost-sensitive learning through the use of PyTorch \citep{paszke2019pytorch}, which calculates cross-entropy loss. 

\vspace{2mm}
\noindent \textbf{Model}
We use an ensemble approach combining two pretrained BERT-based models: BERT \citep{devlin2018bert} and RoBERTa \citep{liu2019roberta}. We take BERT and RoBERTa models independently as baselines. BERT is a \changed{Transformer-based} language representation model that has achieved state-of-the-art performances on numerous NLP tasks. \changed{BERT consists of multiple Transformer encoder layers \cite{vaswani2017attention} with self-attention heads. Its input vector is computed by tokenizing text sequences into wordpieces and feeding them to three embedding layers (token, position, and segment). The following are BERT's pretraining objectives: (1) masked language modeling (predicting input tokens that have been randomly masked) and (2) next sentence prediction (predicting if two input sentences are adjacent to each other).}
Specifically, we use a BERT base (uncased) model from Hugging Face \cite{wolf2020transformers}. \newchanged{It consists of 12 layers of transformers block with a hidden size of 768 and a number of self-attention heads as 12.} In order to use BERT for our classification task, we \changed{fine-tune the model by adding a fully connected linear layer on top of the BERT layers, which maps representations from the final BERT layer to our two classes of interest.} RoBERTa is a variation of BERT with carefully tuned hyperparameters and more extensive training data. \changed{More precisely, RoBERTa was trained on a 160GB text dataset, which is more than ten times larger than BERT's training set. Furthermore, it employs a dynamic masking strategy during training, allowing the model to learn more robust and comprehensive word representations.} We load the RoBERTa base model to a RobertaForSequenceClassification model supported by Hugging Face.

Inspired by \citet{akiti2020contextual}'s work, we fine-tune the comment-level representations obtained from our base models using a Masked Language Model (LM) and unlabeled comment dataset. A primary objective of the LM is to predict what the masked word is given the context words surrounding a mask token. This process is intended to help our language models generalize better on new examples. For this task, we train the Masked LM for 2 epochs with the default hyperparameters.

We initially train four models: one pair of our baseline models (BERT and RoBERTa) with cost-sensitive learning, and another pair without cost-sensitive learning. We then generate an ensemble model by combining the outputs of the baseline models using a weighted-average ensembling method. In the ensemble model, pretrained weights are loaded prior to training. That is, we retrain BERT and RoBERTa simultaneously based on the aggregated predictions. \changed{We use 90 percent of the data for training/validation and 10 percent of the data for testing. We apply a stratified sampling method to maintain a class distribution of the original dataset.}

\begin{table*}[]
\centering
{\small
\begin{tabularx}{\textwidth}{|l|X|X|X|X|X|X|X|X|}
\hline
\multirow{2}{*}{} & \multicolumn{4}{c|}{\textbf{No sampling}}                                  & \multicolumn{4}{c|}{\textbf{Cost-sensitive learning}}                      \\ \cline{2-9} 
                  & Accuracy       & Precision    & Recall       & F1             & Accuracy       & Precision    & Recall       & F1             \\ \hline
BERT              & 83.14          & \textbf{0.80} & 0.57          & 0.76           & 83.00             & 0.63           & 0.62          & 0.77          \\ \hline
RoBERTa           & 83.27          & 0.67        & 0.66         & 0.78          & 82.38          & 0.63          & \textbf{0.73} & \textbf{0.78}          \\ \hline
BERT+RoBERTa      & \textbf{84.08} & 0.69          & \textbf{0.67} & \textbf{0.79} & \textbf{83.05} & \textbf{0.65} & 0.72          & \textbf{0.78} \\ \hline
\end{tabularx}
}
\caption{Averaged results after 5-fold cv for Task 1. Precision and recall for the minority \emph{(support-giving)} class are reported.} 
\label{tab:support_result}
\end{table*}

\vspace{2mm}
\noindent \textbf{Hyperparameters}
For our baselines of Task 1, we finetune BERT base and RoBERTa base models for 3 epochs with a maximum sequence length of 50 and a batch size of 16 for predicting each label separately. We finetune the model with a learning rate of  \(2 \times 1e-5\), a weight decay of \(1 \times 1e-5\), and 30 steps for warm-up. We use Adam optimizer with the max grad norm set to 0.1. \changed{Early stopping is also employed to avoid overfitting on the training set.}

Prior research suggests that adjusting weights based on performance is preferable to simply utilizing the class proportions of the training set \cite{madabushi2020cost, ling2008cost}. Hence, for cost-sensitive classification, we experimented with different sets of weight values (ranging from 1.5 to 10) for the minority class. The best performing weight combination was 1 and 1.5 for the majority and the minority class, respectively. Likewise, weights for the ensemble model were tested thoroughly and set to [0.8, 0.2].

\vspace{2mm}
\noindent \textbf{Evaluation} We compare the performance of the three models finetuned for the supportiveness classification task: BERT base, RoBERTa base, and an ensemble of these two models. We implement 5-fold cross validation to retrieve generalized results of our model. Considering an unequal distribution of training data, we report accuracy, precision (for the minority class), recall (for the minority class), and macro F1. This is the same evaluation metrics adopted by \citet{dadu2020bert}, which ranked the highest at predicting support in the AAAI workshop. As illustrated in Table \ref{tab:support_result}, the ensemble model achieves the best performance regardless of the inclusion of cost-sensitive learning, particularly for accuracy and F1, when compared to our baseline models. Although the cost-sensitive learning method is effective in increasing the recall for the minority class, it sacrifices precision scores and does not enhance the overall performance of the models. 

As the last step, we create an additional ensemble model by picking the best-performing ensemble model from each training technique (cost-sensitive learning vs. regular training). This model is evaluated on unseen test data. We again run multiple experiments with different weight combinations and carefully choose model weights as [0.5, 0.5]. Our proposed model achieves \emph{85\% accuracy, 0.68 precision, 0.74 recall, and 0.8 F1} (see Table \ref{tab:overall_result}). Our final model achieves a slight improvement in recall (0.74 vs. 0.724) and F1 (0.8 vs. 0.774) over the state-of-the-art method \cite{dadu2020bert}.

\subsection{Task 2. Support Type Classification}
\textbf{Method} This task uses the `informational support' and `emotional support' gold labels to detect the type of support offered by content. We train two separate binary classification models using each label, instead of using one multiclass classification model. This will allow us to reuse our proposed model architecture and test its robustness for different purposes. We do not utilize the entire dataset, as Task 1 removes the non-support-giving comments. Thus, we eliminate the negative class of the `support' label. We then apply both training methods (regular training and cost-sensitive training) to models for comparison. \changed{We use 90 percent of the data for training/validation and 10 percent of the data for testing. The stratified sampling method is employed to preserve the original class distribution.}


\begin{table*}[]
\centering
{\small
\begin{tabular}{|c|c|c|c|c|c|c|c|c|}
\hline
\multirow{2}{*}{} & \multicolumn{4}{c|}{\textbf{Emotional (cost-sensitive learning)}}         & \multicolumn{4}{c|}{\textbf{Informational (regular learning)}}             \\ \cline{2-9} 
                  & Accuracy         & Precision   & Recall      & F1            & Accuracy         & Precision      & Recall        & F1            \\ \hline
BERT              & 77.15          & 0.62 & 0.70           & 0.74         & 78.21          & 0.72           & 0.71          & 0.77          \\ \hline
RoBERTa           & 77.06          & 0.62          & \textbf{0.71} & 0.74          & \textbf{79.14} & \textbf{0.73} & \textbf{0.72} & \textbf{0.78} \\ \hline
BERT+RoBERTa      & \textbf{77.28} & \textbf{0.64} & 0.68          & \textbf{0.74} & 79.00    & 0.72 & 0.69          & 0.77 \\ \hline
\end{tabular}
}
\caption{Averaged results after 5-fold cv for Task 2. Precision and recall for the minority \emph{(support-giving)} class are reported.} 
\label{tab:task2_result}
\end{table*}

\begin{table*}[]
\centering
{\small
\begin{tabular}{|c|c|c|c|c|}
\hline
\textbf{Label}                      & \textbf{Accuracy} & \textbf{Precision} & \textbf{Recall} & \textbf{F1}   \\ \hline
Support               & 84.73       & 0.68      & 0.74   & 0.80  \\ \hline
Informational support & 78.94    & 0.74      & 0.73      & 0.78 \\ \hline
Emotional support     & 80.80     & 0.69      & 0.71    & 0.78 \\ \hline
\end{tabular}
}
\caption{Final outcomes of proposed models. Precision and recall for the minority \emph{(support-giving)} class are reported.} 
\label{tab:overall_result}
\end{table*}

\vspace{2mm}
\noindent \textbf{Model} 
We experiment with the same three model architectures (BERT, RoBERTa, BERT + RoBERTa) presented in Task 1. This time, however, we do not load the pretrained weights on the ensemble model, as a gap between training errors and validation errors has begun to grow. This may be due to our relatively small dataset size (n=3,226); training the model iteratively using the small data is known to induce overfitting.

\vspace{2mm}
\noindent \textbf{Evaluation} We perform 5-fold cross validation on each model to compare their performance. We utilize the same evaluation metrics (accuracy, precision (for the minority class), recall (for the minority class), macro F1) implemented in Task 1. As a result of comparison (regular training vs. cost-sensitive training), cost-sensitive learning is selected for the emotional support classifier, whereas we employ a regular training process for the `informational support' label. Table \ref{tab:task2_result} shows their results.

We find that the the ensemble model's performance (specifically, accuracy and precision) is slightly better than the performance of the baseline models with respect to the `emotional support' label. On the other hand, for informational support classification, RoBERTa outperforms all other models with an increase of 1 percent in precision, recall, and F1 score.

Our final model is the ensemble of two best performing models across 5 folds. For emotional support labeling, we select two ensemble-based models. For informational support labeling, we choose two finetuned RoBERTa models. To perform a weighted-average prediction, we compare the performance using several weight combinations that sum up to 1.0 and finally select [0.5,0.5] and [0.6, 0.4] for two ensemble models. Table \ref{tab:overall_result} demonstrates the performance of all three models on the test dataset, indicating the effectiveness of ensembling techniques.

\begin{table}[]
\centering
{\small
\begin{tabular}{|c|c|}
\hline
\textbf{Categories}         & \textbf{N (\%)}      \\ \hline
Age                & 338 (14.09) \\ \hline
Race/ethnicity     & 10 (0.42)   \\ \hline
Gender             & 141 (5.88)  \\ \hline
Marital status     & 399 (16.63) \\ \hline
Education          & 266 (11.09) \\ \hline
Employment status  & 475 (19.8)  \\ \hline
Location           & 386 (16.09) \\ \hline
Health information & 566 (23.59) \\ \hline
Phone number       & 0 (0)       \\ \hline
Email address      & 0 (0)       \\ \hline
SSN                & 0 (0)       \\ \hline
No disclosure      & 859 (35.81)        \\ \hline
\end{tabular}
}
\caption{Statistics of personal information categories} 
\label{tab:person_info}
\end{table}

\definecolor{bblue}{HTML}{4F81BD}
\definecolor{rred}{HTML}{C0504D}
\definecolor{ggreen}{HTML}{9BBB59}
\definecolor{ppurple}{HTML}{9F4C7C}

\subsection{Human Evaluation of Model Performance}
Our proposed model is trained on \#OffMyChest Reddit corpus, and we use this approach to label our COVID-19 support dataset without the help of human annotators. Yet, it is uncertain if the classifier can generalize well to the new setting. Thus, two researchers re-labeled a random subset of annotated posts (200) to evaluate the trustworthiness of the labels generated from our AI-driven model. To ensure the integrity of annotation for every category, the number of instances in each category of the subset was equally distributed. Human annotators were given a detailed annotation instruction provided by the AffCon Workshop\footnote{Their labeling instructions can be found in this Github repository: \url{https://github.com/kj2013/claff-offmychest}.} at AAAI 2019. We then calculated Fleiss’ Kappa amongst two annotators and labels created by our model. The results of calculated Kappa scores for each category are as follows: support (0.78), informational support (0.72), and emotional support (0.74). Given the observed moderate agreement rate, we conclude that our machine-generated labels are reliable enough to continue the analyses.


\section{Results} \label{sec:findings}

\begin{figure*}[h!]
\centering
\begin{adjustbox}{max width=\textwidth}
\begin{tikzpicture}
\begin{axis}[
ybar=0pt,
ymax=0.3,
width=\textwidth,
major x tick style = transparent,
height=2.5in,
bar width=4mm,
enlarge y limits={0.05,upper},
enlarge x limits=0.05,
ymajorgrids = true,
legend cell align=left,
legend style={at={(0.5,-0.4)},
anchor=north,legend columns=-1, column sep=1ex },
ylabel={A proportion of posts},
symbolic x coords={Age,Gender, Marital\\Status,Education,Employment\\Status,Location, Health\\Information},
xtick=data,
x tick label style={rotate=45, align=left, font=\small},
style={font=\small}
]

\addplot[style={black,fill=bblue,mark=none}]
    coordinates {(Age,0.1) (Gender, 0.04) (Marital\\Status,0.17) (Education,0.06) (Employment\\Status,0.16) (Location,0.14) (Health\\Information,0.25)};

\addplot [style={black,fill=rred,mark=none}]
    coordinates {(Age,0.14) (Gender, 0.06) (Marital\\Status,0.17) (Education,0.12) (Employment\\Status,0.2) (Location,0.15) (Health\\Information,0.23)};

\addplot [style={black,fill=ggreen,mark=none}]
    coordinates {(Age,0.17) (Gender, 0.08) (Marital\\Status,0.18) (Education,0.14) (Employment\\Status,0.22) (Location,0.2) (Health\\Information,0.26)};

\legend{Informational, Emotional, Both}
\end{axis}
\end{tikzpicture}
\end{adjustbox}
\caption{Distribution of personal information categories in requested social support}
\label{fig:personal_distribution}
\end{figure*}
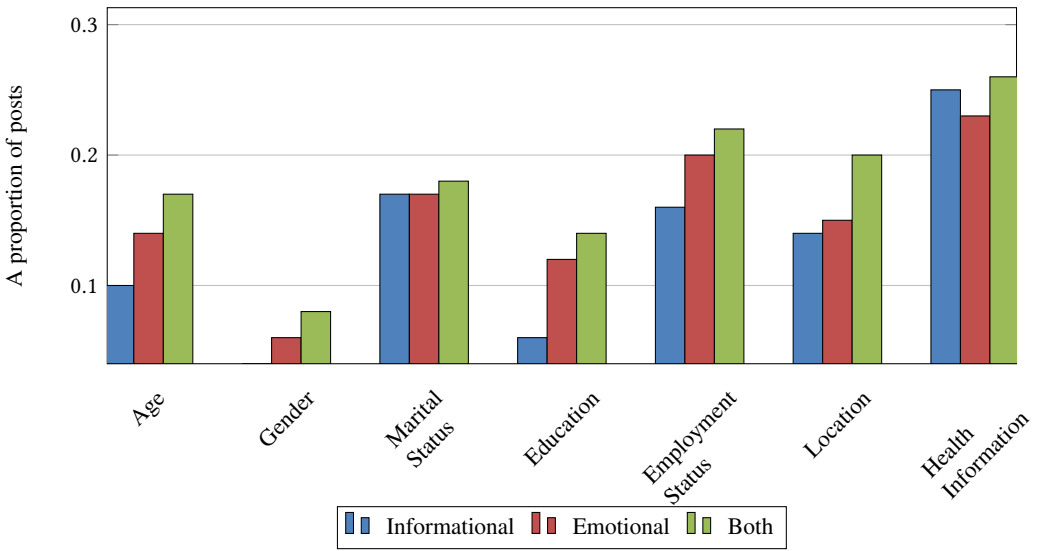

Following, we discuss statistical analyses supporting our three primary research questions. In support of the first research question (RQ1), we analyzed the patterns of self-disclosure within three social support groups (informational support vs. emotional support vs. both) for each personal information category through a Chi-Square test. For RQ2, linear regression models were employed to study the relationship between voluntary information sharing practices and receptions of requested support and user engagement. Five user engagement features are submission scores, the number of comments and unique commenters, the average length of comments, and their sentiment levels. Finally, turning to the third research question (RQ3), we focused on multiple linguistic and topical attributes of postings with an element of self-disclosure and examine their roles in social \changed{support and user involvement} acquisition using stepwise regression models. All statistical tests were performed using R, an open‐source statistical software package. \changed{Data was screened for multicollinearity, homoscedasticity, and normality assumptions.} 

\subsection{Annotation Analysis} 
Here we present annotation results of COVID19\_support dataset for self-disclosure and social support tagging. All labels are obtained using approaches described in Sections \ref{sec:manual} and \ref{sec:automated}.

\vspace{2mm}
\noindent \textbf{Self-Disclosure} Table \ref{tab:person_info} gives the personal information category breakdown of the collected posts. Of 2,399 posts, 1,540 posts (64.19\%) contain an element of objective self-disclosure. Users' public discourse about health information (23.59\%) was prevalent in contrast to other categories. Furthermore, users frequently shared personal data with respect to their employment status (19.8\%), marital status (16.63\%), and location (16.09\%). The dissemination of classic types of personally identifiable information (PHI) including phone number, email address, and social security number is not identified since one of the `r/COVID19\_support' rules explicitly mentions that users should not disclose it. Meanwhile, 859 (35.81\%) are determined not to contain any form of objective self-disclosure.

\begin{table}[]
{\small
\begin{tabular}{|c|c|c|}
\hline
\textbf{Categories}            & \textbf{Posts} & \textbf{Comments} \\ \hline
Support               & 2,330 & 10,917   \\ \hline
Informational support & 953   & 4,979    \\ \hline
Emotional support     & 1,992 & 5,104    \\ \hline
\end{tabular}
}
\caption{Statistics of labeled posts and comments for social support exchange}
\label{tab:support_stat}
\end{table}

\vspace{2mm}
\noindent \textbf{Social Support} Statistics of our labeled COVID19\_support dataset are presented in Table \ref{tab:support_stat}. We find that, of 2,339 posts under the `Support' flair, 2,331 requested either informational, emotional, or both types of support. More precisely, 39.76\% seek informational assistance, whereas 85.16\% request emotional support. Among 29,851 comments that replied to these posts, 4,979 (16.68\%) provide informational help, and 5,104 (17.1\%) deliver sentimental assistance. Approximately one-third of the comments offer any form of support for the posters.

\subsection{RQ1. Self-Disclosure and Support-Seeking \newchanged{during COVID-19}}

\noindent Of the 2,399 posts in the r/COVID19\_support group, 2,331 requested either informational, emotional, or both types of support. The 68 posts that did not seek any support were not considered in the analysis for RQ1. Additionally, phone number, email address, SSN, and race categories for self-disclosure were excluded from the analysis as they have fewer than 10 instances. Figure \ref{fig:personal_distribution} shows that users generally shared their personal information most frequently when seeking both types of support. On the contrary, informational support seekers were less associated with public self-disclosure, with the exception of health information. Subsequently, we run a Chi-Square Test of Independence to test the significance of differences within three groups for each category. As displayed in Table \ref{tab:chi2-results}, the age ($p = 0.006$), location ($p = 0.038$), and education ($p = 0.002$) categories are found to be significant with p-values less than 0.05. P-values for the remaining categories are 0.127 (race), 0.073 (marital status), 0.059 (gender), and 0.352 (health information). To further uncover which support-seeking types are different, post-hoc pairwise testing was conducted. For the age category, we tested the rate of self-disclosure between informational and emotional support (9.73\% vs. 14.15\%, $p = 0.039$) and the difference between informational and both types of support (9.73\% vs. 17.26\%, $p = 0.002$). For location, the difference between emotional and both support (14.45\% vs. 19.54\%) was determined to be significant ($p = 0.023$). For the education category, the difference between informational and emotional (5.9\% vs. 11.76\%, p = 0.002) and the difference between informational and both (5.9\% vs. 13.52\%, $p = 0.0004$) are significant.

\begin{table*}[t]
\begin{center}
{\small
\begin{tabular}{|c|c|c|c|c|c|c|c|}
\hline
\multirow{2}{*}{}  & \multicolumn{2}{c|}{\textbf{Informational}} & \multicolumn{2}{c|}{\textbf{Emotional}} & \multicolumn{2}{c|}{\textbf{Both}} & \multirow{2}{*}{\emph{p}} \\ \cline{2-7}
                   & N              & \%                & N            & \%              & N          & \%           &                               \\ \hline
Age                & 33             & 9.73            & 195          & 14.15          & 106        & 17.26       & 0.006**                    \\ \hline
Gender             & 14             & 4.13            & 78           & 5.66          & 48         & 7.82       & 0.051                       \\ \hline
Marital status            & 59             & 17.4            & 228          & 16.55          & 110        & 17.92       & 0.739                        \\ \hline
Education          & 20             & 5.9            & 162          & 11.76           & 83         & 13.52        & 0.002**                    \\ \hline
Employment status       & 54             & 15.93            & 279          & 20.25          & 138        & 22.48       & 0.055                       \\ \hline

Location           & 49             & 14.45            & 211          & 15.31          & 120        & 19.54       & 0.038*                     \\ \hline

Health information & 85             & 25.07            & 316          & 22.93          & 158        & 25.73       & 0.352                        \\ \hline

\end{tabular}
}
 \begin{tablenotes}
      \small
      \item Signif. codes: ‘***’ 0.001 ‘**’ 0.01 ‘*’ 0.05
\end{tablenotes}
\caption{Chi-Square results for RQ1}
\label{tab:chi2-results}
\end{center}
\end{table*}

\subsection{RQ2. Self-Disclosure and \changed{the Receipt of Social Support and User Engagement} \newchanged{during COVID-19}}
\noindent \textbf{Self-Disclosure and Support-Receiving} Of 2,399 posts, 2,328 posts were followed by at least one responsive comment. The remaining 71 posts that did not receive any response were removed from the analyses. We quantify the amount of self-disclosure (independent variable) by aggregating the total number of personal information categories shared in each post. The dependent variable is calculated as the percentage of comments that successfully offer the same type of social support that the poster is pursuing. For instance, if the poster seeks informational support and there are 4 informational support-providing comments out of 8 comments, the calculated value will be 0.5. As both the independent variable and the dependent variable are continuous, we perform a linear regression for each support type (informational vs. emotional) independently.

In terms of informational support, our results indicate that the amount of self-disclosure in a post and the receipt of matching support are statistically correlated ($p = 0.007$). Coefficients of the model indicate that for every one-unit increase in the summation of personal information categories disclosed in a post, the percentage of matching received support was greater by 4.61\%. For emotional support, on the other hand, our predictor variable was found to be statistically insignificant ($p = 0.155$).

\begin{table*}[]
{\small
\begin{tabular}{|c|cc|cc|}
\hline
\multirow{2}{*}{}      & \multicolumn{2}{c|}{\textbf{Informational}}             & \multicolumn{2}{c|}{\textbf{Emotional}}                 \\ \cline{2-5} 
                       & \multicolumn{1}{c|}{Coefficient} & \emph{p}           & \multicolumn{1}{c|}{Coefficient} & \emph{p}           \\ \hline
Submission scores      & \multicolumn{1}{c|}{1.71}    & 0.04*     & \multicolumn{1}{c|}{-3.29}  & 0.001**  \\ \hline
Comment count          & \multicolumn{1}{c|}{-0.004}     &  0.65       & \multicolumn{1}{c|}{-0.11}    & <2e-16***   \\ \hline
Unique commenter count & \multicolumn{1}{c|}{0.01}     & 0.32       & \multicolumn{1}{c|}{-0.09}    & <2e-16*** \\ \hline
Comment length         & \multicolumn{1}{c|}{44.35}     & 8.92e-08*** & \multicolumn{1}{c|}{37.99}     & 4.22e-12*** \\ \hline
Sentiment scores       & \multicolumn{1}{c|}{0.02}     & 0.02*     & \multicolumn{1}{c|}{0.017}     & 0.01**  \\ \hline
\end{tabular}
}
 \begin{tablenotes}
      \small
      \item Signif. codes: ‘***’ 0.001 ‘**’ 0.01 ‘*’ 0.05
\end{tablenotes}
\caption{Regression results for RQ2 (user engagement)}
\label{tab:regression_userfeedback}
\end{table*}

\vspace{2mm}
\noindent \textbf{Self-Disclosure and User Engagement}
Measurements for user engagement include submission scores, the number of corresponding comments and unique commenters, comment lengths, and sentiment scores. Descriptions of how we quantify these features are included below:
\begin{itemize}
\item \emph{submission scores}: 
We calculate the submission score of a post by subtracting the number of downvotes from the number of upvotes.
\item \emph{number of comments}: A total number of comments associated with each post, regardless of commenter ID, is calculated. 
\item \emph{number of unique commenters}: We count the number of non-overlapping commenters who reacted to a post.
\item \emph{length of comments}: We calculate the average number of characters included in comments.
\item \emph{sentiment scores}: \newchanged{We use the VADER algorithm \citep{hutto2014vader} to retrieve a compound polarity score, where an overall sentiment level is normalized between -1 (extremely negative) and 1 (extremely positive).  For each post, we aggregate compound scores of all associated comments and then average them as a final score.}

\end{itemize}


\noindent For statistical testing, we consider self-disclosure as an independent variable and five user engagement features as dependent variables. \newchanged{The rate of self-disclosure is quantified in a manner equivalent to the previous analyses.} Since there are two support types (informational, and emotional) and five feedback features (submission scores, number of comments, number of unique commenters, length of comments, and sentiment scores), we run eight separate linear regression models in total. While linear regression models are employed for dependent variables like submission scores, sentiment scores, and average comment length, we adopt  Poisson regression models for the remaining variables because they are discrete count variables. 

Results of our regression analysis are in Table \ref{tab:regression_userfeedback}. We find that, for both support types, there exist statistically significant correlations between the degree of self-disclosure in a post and three user engagement features, including submission scores, comment length, and sentiment scores. Taking into account the positive coefficient values of the model, the results suggest that, when seeking informational support, as the rate of self-disclosure increases, submission scores, comment length, and sentiment scores tend to increase. The observed trends remain consistent for emotional support seekers, except for submission scores; their voluntary discourse is negatively coupled with submission scores. Similarly, emotional support seekers' self-disclosure is not related to a higher amount of comments or commenters. For informational support, on the other hand, both predictor variables were found to be statistically insignificant.

\begin{table}[]
{\small
\begin{tabular}{|c|c|c|}
\hline
\textbf{Name} & \textbf{Keywords}                            \\ \hline
Topic 1       & room, clean, water, bathroom, wash, hotel                      \\ \hline
Topic 2       & covid, anxiety, feel, scar, symptoms, like                  \\ \hline
Topic 3       & feel, know, time, people, want, friends                      \\ \hline
Topic 4       & taste, skills, experts, discourage, anybody                \\ \hline
Topic 5       & test, positive, mom, dad, result, work                      \\ \hline
Topic 6       & stomach, surface, shortness\_breath, crush                    \\ \hline
Topic 7       & winter, spring, summer, seasonal, surge                    \\ \hline
Topic 8       & family, people, live, take, home                           \\ \hline
Topic 9       & job, work, school, college, online, class                    \\ \hline
Topic 10       & emotional, abusive, serious, boyfriends, abuse                 \\ \hline
Topic 11      & suggestions, body\_ache, vitamin, cold                     \\ \hline
Topic 12      & mask, wear, outside, walk, distance                        \\ \hline
Topic 13      & party, huge, isolation, lead                            \\ \hline
Topic 14      & luck, mood, wife, permanent, symptomatic                 \\ \hline
Topic 15      & agree, usa, regret, confirm, forward                       \\ \hline
\end{tabular}
}
\caption{Frequently discussed topics of support-seeking posts with self-disclosure}
\label{tab:topic_des}
\end{table}

\begin{table}[]
{\small
\begin{tabular}{|ccc|ccc|}
\hline
\multicolumn{3}{|c|}{\textbf{Informational}}                                                                                                                & \multicolumn{3}{c|}{\textbf{Emotional}}                                                                                                    \\ \hline
\multicolumn{1}{|c|}{Variable}                                                       & \multicolumn{1}{c|}{Coefficient}            & \emph{p}                      & \multicolumn{1}{c|}{Variable}                                                             & \multicolumn{1}{c|}{Coefficient} & \emph{p}           \\ \hline
\multicolumn{1}{|c|}{\begin{tabular}[c]{@{}c@{}}Education\end{tabular}} & \multicolumn{1}{c|}{0.06}                   & 0.066                  & \multicolumn{1}{c|}{\begin{tabular}[c]{@{}c@{}}Age\end{tabular}}             & \multicolumn{1}{c|}{0.024}       & 0.108       \\ \hline
\multicolumn{1}{|c|}{Topic 2}                                                        & \multicolumn{1}{c|}{0.187}                  & 0.081                  & \multicolumn{1}{c|}{\textbf{\begin{tabular}[c]{@{}c@{}}Positive \\ sentiment\end{tabular}}} & \multicolumn{1}{c|}{-0.767}      & 0.025*      \\ \hline
\multicolumn{1}{|c|}{\textbf{Topic 5}}                                               & \multicolumn{1}{c|}{0.277}                  & 0.035*                 & \multicolumn{1}{c|}{\textbf{\begin{tabular}[c]{@{}c@{}}Neutral\\ sentiment\end{tabular}}}   & \multicolumn{1}{c|}{-0.754}      & 0.0007***   \\ \hline
\multicolumn{1}{|c|}{\textbf{Topic 8}}                                               & \multicolumn{1}{c|}{0.322}                  & 0.004**                & \multicolumn{1}{c|}{\begin{tabular}[c]{@{}c@{}}Compound \\ sentiment\end{tabular}}          & \multicolumn{1}{c|}{-0.027}      & 0.058       \\ \hline
\multicolumn{1}{|c|}{Topic 11}                                                       & \multicolumn{1}{c|}{1.815}                  & 0.082                  & \multicolumn{1}{c|}{\textbf{Readability}}                                                 & \multicolumn{1}{c|}{0.002}       & 0.015*      \\ \hline
\multicolumn{1}{|c|}{\multirow{7}{*}{Topic 12}}                                      & \multicolumn{1}{c|}{\multirow{7}{*}{0.384}} & \multirow{7}{*}{0.061} & \multicolumn{1}{c|}{\textbf{First-person count}}                                                  & \multicolumn{1}{c|}{0.001}       & 0.002**     \\ \cline{4-6} 
\multicolumn{1}{|c|}{}                                                               & \multicolumn{1}{c|}{}                       &                        & \multicolumn{1}{c|}{Topic 2}                                                              & \multicolumn{1}{c|}{0.427}       & 0.003**     \\ \cline{4-6} 
\multicolumn{1}{|c|}{}                                                               & \multicolumn{1}{c|}{}                       &                        & \multicolumn{1}{c|}{\textbf{Topic 3}}                                                     & \multicolumn{1}{c|}{0.653}       & 5.44e-07*** \\ \cline{4-6} 
\multicolumn{1}{|c|}{}                                                               & \multicolumn{1}{c|}{}                       &                        & \multicolumn{1}{c|}{\textbf{Topic 5}}                                                     & \multicolumn{1}{c|}{0.792}       & 1.66e-07*** \\ \cline{4-6} 
\multicolumn{1}{|c|}{}                                                               & \multicolumn{1}{c|}{}                       &                        & \multicolumn{1}{c|}{\textbf{Topic 8}}                                                     & \multicolumn{1}{c|}{0.457}       & 0.002**     \\ \cline{4-6} 
\multicolumn{1}{|c|}{}                                                               & \multicolumn{1}{c|}{}                       &                        & \multicolumn{1}{c|}{\textbf{Topic 9}}                                                     & \multicolumn{1}{c|}{0.689}       & 7.52e-07*** \\ \cline{4-6} 
\multicolumn{1}{|c|}{}                                                               & \multicolumn{1}{c|}{}                       &                        & \multicolumn{1}{c|}{\textbf{Topic 10}}                                                     & \multicolumn{1}{c|}{3.476}       & 0.034*      \\ \cline{4-6} 
\hline
\end{tabular}
}
 \begin{tablenotes}
      \small
      \item Signif. codes: ‘***’ 0.001 ‘**’ 0.01 ‘*’ 0.05
\end{tablenotes}
\caption{Regression results for RQ3 (support receiving)}
\label{tab:rq3_regression_1}
\end{table}

\changed{\subsection{RQ3. Linguistic Markers of Self-Disclosure and the Receipt of Social Support and User Engagement \newchanged{during COVID-19}}}

Turning to RQ3, we examine whether textual characteristics of self-disclosing posts are predictive of the acquisition of the requested social support type. Here we utilize six post attributes as predictor variables: personal information categories, post length, count of first-person pronouns, readability scores, sentiment scores (negative, neutral, positive, and compound), and topic distributions. Detailed explanations for each variable are as follows:
\begin{itemize}
\item \emph{personal information categories}: we rely on our annotation results of self-disclosure types. Since none of the posts share the authors' phone number, email address, or SSN, this leads to 8 binary variables in total. 
\item \emph{post length}: We count the number of characters within a post.
\item \emph{first-person pronouns}: A total number of first-person pronouns such as `I', `my', `me', `mine', and `myself' included in a post is calculated. For text pre-processing, we first expand contractions (e.g., I'm -> I am) using a Python package named `contractions'\footnote{https://github.com/kootenpv/contractions} and lower-case sentences.
\item \emph{readability scores}: Readability of a post is measured with the Flesch Reading Ease (FRE) metric, which takes the number of words per sentence and the number of syllables per word into account \citep{flesch1948new}. the FRE gives a text a score between 1 and 100, with 100 being the highest readability score. This metric has been widely adopted by existing literature \citep{chen2020linguistic, oztok2013exploring} and is calculated as: 

\begin{align*}
FRE=206.835-1.015*\frac{total\:words}{total\:sentences}-  84.6*\frac{total\:syllables}{total\:words}
\end{align*} 
\vskip 0.2in

\item \emph{sentiment scores}: We use the VADER algorithm to measure four sentiment components (positive, negative, neutral, and compound) of a post.
\item \emph{topic distributions}: We employ the Latent Dirichlet Allocation (LDA) algorithm \citep{blei2003latent}, one of the most popular probabilistic topic models that automatically extracts latent topics when a set of documents is given. We use the Gensim\footnote{https://radimrehurek.com/gensim/} package to implement the LDA. To find the best number of topics \emph{N}, we compare topic coherence measures resulting from $\emph{N} \subset \{10,15,20,25,30,35,40,45\}$. Finally, \emph{N} is set to 15 (topic coherence = 0.481). Identified topics are illustrated in Table \ref{tab:topic_des}.
\end{itemize}

\vspace{2mm}
\noindent \textbf{Post Attributes and Support-Receiving} 
\changed{We have 30 independent variables from textual features of self-disclosing posts.} Equivalent to RQ2, the percentage of comments giving the same form of social support that the poster is pursuing is used to determine the dependent variable. We fit a multivariate stepwise regression model to select a subset of features that yielded the best prediction for the receipt of appropriate support. Of 2,328 posts that received at least one comment, 1,495 posts contain personal data, and hence the remaining 833 are ignored from statistical analyses. \changed{Since we have two support types (informational and emotional support) sought by post authors, we differentiate them and build two independent models.}

\newchanged{Table \ref{tab:rq3_regression_1} shows results of regression models with selected features. For informational support, the variables \emph{education information disclosure, topic 2, topic 5, topic 8, topic 11, and topic 11} are chosen. When we run a new regression model with these features, topic 5 and topic 8 are statistically significant ($p = 0.035; p = 0.004$). Specifically, taking into account their positive coefficient values, posts about family members or their COVID-19 test results are positively related to appropriate support acquisition when the post authors need informational support.
For emotional support, we find that 12 predictor variables including age information disclosure, positive and neutral sentiments, readability scores, the inclusion of the first-person pronouns, and 5 topics relevant to general health concerns to emotional burdens are most effective in predicting the receipt of desired emotional support. Specifically, the higher levels of positive or neutral emotions within the post, the lower the amount of appropriate emotional support seeking individuals obtain. On the contrary, the use of first-person pronouns and post readability are found to be positively correlated to the reciprocity of emotional support.}

\begin{table}[]
{\small
\begin{tabular}{|ccc|ccc|}
\hline
\multicolumn{3}{|c|}{\textbf{Informational}}                                                                                                                          & \multicolumn{3}{c|}{\textbf{Emotional}}                                                                                                     \\ \hline
\multicolumn{1}{|c|}{Variable}                                                               & \multicolumn{1}{c|}{Coefficient}            & \emph{p}                        & \multicolumn{1}{c|}{Variable}                                                       & \multicolumn{1}{c|}{Coefficient} & \emph{p}                  \\ \hline
\multicolumn{1}{|c|}{\textbf{Age}}                                                           & \multicolumn{1}{c|}{0.079}                  & 0.006**                  & \multicolumn{1}{c|}{\textbf{Age}}                                                   & \multicolumn{1}{c|}{0.096}       & 5.01e-10***        \\ \hline
\multicolumn{1}{|c|}{\textbf{Employment}}                                                    & \multicolumn{1}{c|}{0.049}                  & 0.028*                   & \multicolumn{1}{c|}{\textbf{Employment}}                                            & \multicolumn{1}{c|}{-0.171}      & \textless 2e-16*** \\ \hline
\multicolumn{1}{|c|}{\textbf{Race}}                                                          & \multicolumn{1}{c|}{-0.287}                 & 0.005**                  & \multicolumn{1}{c|}{\textbf{Race}}                                                  & \multicolumn{1}{c|}{-0.338}      & 0.0003***          \\ \hline
\multicolumn{1}{|c|}{\textbf{Location}}                                                      & \multicolumn{1}{c|}{-0.113}                 & 2.94e-06 ***             & \multicolumn{1}{c|}{\textbf{Location}}                                              & \multicolumn{1}{c|}{-0.102}      & 1.08e-10***        \\ \hline
\multicolumn{1}{|c|}{\textbf{Education}}                                                     & \multicolumn{1}{c|}{-0.099}                 & 0.0008***                & \multicolumn{1}{c|}{\textbf{Education}}                                             & \multicolumn{1}{c|}{-0.202}      & \textless 2e-16*** \\ \hline
\multicolumn{1}{|c|}{\textbf{Marital status}}                                                & \multicolumn{1}{c|}{-0.056}                 & 0.027*                   & \multicolumn{1}{c|}{\textbf{Health information}}                                           & \multicolumn{1}{c|}{-0.05}       & 0.002**            \\ \hline
\multicolumn{1}{|c|}{Gender}                                                                 & \multicolumn{1}{c|}{-0.076}                 & 0.051                    & \multicolumn{1}{c|}{\textbf{Marital status}}                                        & \multicolumn{1}{c|}{-0.092}      & 1.42e-08***        \\ \hline
\multicolumn{1}{|c|}{\textbf{\begin{tabular}[c]{@{}c@{}}Positive \\ sentiment\end{tabular}}} & \multicolumn{1}{c|}{-153.5}                 & 5.36e-15***              & \multicolumn{1}{c|}{\begin{tabular}[c]{@{}c@{}}Positive\\ sentiment\end{tabular}}  & \multicolumn{1}{c|}{23.82}       & 0.063              \\ \hline
\multicolumn{1}{|c|}{\textbf{\begin{tabular}[c]{@{}c@{}}Neutral\\ sentiment\end{tabular}}}   & \multicolumn{1}{c|}{-153.6}                 & 5.26e-15***              & \multicolumn{1}{c|}{\begin{tabular}[c]{@{}c@{}}Neutral \\ sentiment\end{tabular}}  & \multicolumn{1}{c|}{23.58}       & 0.066              \\ \hline
\multicolumn{1}{|c|}{\textbf{\begin{tabular}[c]{@{}c@{}}Negative \\ sentiment\end{tabular}}} & \multicolumn{1}{c|}{-150}                   & 2.21e-14***              & \multicolumn{1}{c|}{\begin{tabular}[c]{@{}c@{}}Negative \\ sentiment\end{tabular}} & \multicolumn{1}{c|}{23.77}       & 0.0637             \\ \hline
\multicolumn{1}{|c|}{\textbf{\begin{tabular}[c]{@{}c@{}}Compound\\ sentiment\end{tabular}}}  & \multicolumn{1}{c|}{-0.138}                 & 7.49e-10***              & \multicolumn{1}{c|}{\textbf{Post length}}                                           & \multicolumn{1}{c|}{8.480e-05}   & 5.03e-11***        \\ \hline
\multicolumn{1}{|c|}{\textbf{Post length}}                                                   & \multicolumn{1}{c|}{6.524e-05}              & 0.002**                  & \multicolumn{1}{c|}{\textbf{Readability}}                                           & \multicolumn{1}{c|}{0.011}       & \textless 2e-16*** \\ \hline
\multicolumn{1}{|c|}{\textbf{First-person count}}                                            & \multicolumn{1}{c|}{-0.012}                 & \textless 2e-16***       & \multicolumn{1}{c|}{\textbf{First-person count}}                                    & \multicolumn{1}{c|}{0.008}       & \textless 2e-16*** \\ \hline
\multicolumn{1}{|c|}{\textbf{Topic 1}}                                                       & \multicolumn{1}{c|}{-4.643}                 & 6.01e-11***              & \multicolumn{1}{c|}{\textbf{Topic 1}}                                               & \multicolumn{1}{c|}{-6.66}       & \textless 2e-16*** \\ \hline
\multicolumn{1}{|c|}{\textbf{Topic 2}}                                                       & \multicolumn{1}{c|}{-1.671}                 & \textless 2e-16***       & \multicolumn{1}{c|}{\textbf{Topic 2}}                                               & \multicolumn{1}{c|}{-5.099}      & \textless 2e-16*** \\ \hline
\multicolumn{1}{|c|}{\textbf{Topic 5}}                                                       & \multicolumn{1}{c|}{-1.468}                 & \textless 2e-16***       & \multicolumn{1}{c|}{\textbf{Topic 3}}                                               & \multicolumn{1}{c|}{-2.897}      & 2.18e-07***        \\ \hline
\multicolumn{1}{|c|}{\textbf{Topic 7}}                                                       & \multicolumn{1}{c|}{9.225}                  & \textless 2e-16***       & \multicolumn{1}{c|}{\textbf{Topic 4}}                                               & \multicolumn{1}{c|}{-9.673}      & 2.52e-08***        \\ \hline
\multicolumn{1}{|c|}{\textbf{Topic 8}}                                                       & \multicolumn{1}{c|}{-0.533}                 & 3.59e-07***              & \multicolumn{1}{c|}{\textbf{Topic 5}}                                               & \multicolumn{1}{c|}{-4.426}      & 1.57e-15***        \\ \hline
\multicolumn{1}{|c|}{\textbf{Topic 11}}                                                      & \multicolumn{1}{c|}{-10.86}                 & \textless 2e-16***       & \multicolumn{1}{c|}{\textbf{Topic 6}}                                               & \multicolumn{1}{c|}{-4.408}      & 0.0001***          \\ \hline
\multicolumn{1}{|c|}{\textbf{Topic 12}}                                                      & \multicolumn{1}{c|}{-1.174}                 & 5.54e-08***              & \multicolumn{1}{c|}{\textbf{Topic 8}}                                               & \multicolumn{1}{c|}{-3.383}      & 1.14e-09***        \\ \hline
\multicolumn{1}{|c|}{\textbf{Topic 13}}                                                      & \multicolumn{1}{c|}{-4.403}                 & 1.56e-07***              & \multicolumn{1}{c|}{\textbf{Topic 9}}                                               & \multicolumn{1}{c|}{-3.398}      & 2.26e-09***        \\ \hline
\multicolumn{1}{|c|}{Topic 14}                                                               & \multicolumn{1}{c|}{-3.18}                  & 0.157                    & \multicolumn{1}{c|}{\textbf{Topic 10}}                                               & \multicolumn{1}{c|}{-6.064}      & 0.0004***          \\ \hline
\multicolumn{1}{|c|}{\multirow{5}{*}{\textbf{Topic 15}}}                                     & \multicolumn{1}{c|}{\multirow{5}{*}{4.317}} & \multirow{5}{*}{0.002**} & \multicolumn{1}{c|}{\textbf{Topic 11}}                                              & \multicolumn{1}{c|}{-4.662}      & 1.05e-06***        \\ \cline{4-6} 
\multicolumn{1}{|c|}{}                                                                       & \multicolumn{1}{c|}{}                       &                          & \multicolumn{1}{c|}{\textbf{Topic 12}}                                              & \multicolumn{1}{c|}{-3.470}      & 5.47e-10***        \\ \cline{4-6} 
\multicolumn{1}{|c|}{}                                                                       & \multicolumn{1}{c|}{}                       &                          & \multicolumn{1}{c|}{\textbf{Topic 13}}                                              & \multicolumn{1}{c|}{-6.908}      & \textless 2e-16*** \\ \cline{4-6} 
\multicolumn{1}{|c|}{}                                                                       & \multicolumn{1}{c|}{}                       &                          & \multicolumn{1}{c|}{\textbf{Topic 14}}                                              & \multicolumn{1}{c|}{-0.117}      & 4.33e-13***        \\ \cline{4-6} 
\multicolumn{1}{|c|}{}                                                                       & \multicolumn{1}{c|}{}                       &                          & \multicolumn{1}{c|}{\textbf{Topic 15}}                                              & \multicolumn{1}{c|}{4.725}       & 4.96e-07***        \\ \hline
\end{tabular}
}
\begin{tablenotes}
      \small
      \item Signif. codes: ‘***’ 0.001 ‘**’ 0.01 ‘*’ 0.05
\end{tablenotes}
\caption{Regression results for RQ3 (comment \& commenter count). \changed{Variables that were not useful in prediction based on stepwise regression were omitted.}}
\label{tab:rq3_regression_comment}
\end{table}

\begin{table}[]
{\small
\begin{tabular}{|ccc|ccc|}
\hline
\multicolumn{3}{|c|}{\textbf{Informational}}                                                                                                                        & \multicolumn{3}{c|}{\textbf{Emotional}}                                                                                          \\ \hline
\multicolumn{1}{|c|}{Variable}                                                              & \multicolumn{1}{c|}{Coefficient}            & \emph{p}                       & \multicolumn{1}{c|}{Variable}                                                   & \multicolumn{1}{c|}{Coefficient} & \emph{p}           \\ \hline
\multicolumn{1}{|c|}{Employment}                                                            & \multicolumn{1}{c|}{4.159}                  & 0.136                   & \multicolumn{1}{c|}{Employment}                                                 & \multicolumn{1}{c|}{23.59}       & 0.074       \\ \hline
\multicolumn{1}{|c|}{Location}                                                              & \multicolumn{1}{c|}{-4.924}                 & 0.079                   & \multicolumn{1}{c|}{Location}                                                   & \multicolumn{1}{c|}{-4.728}      & 0.107       \\ \hline
\multicolumn{1}{|c|}{\textbf{\begin{tabular}[c]{@{}c@{}}Positive\\ sentiment\end{tabular}}} & \multicolumn{1}{c|}{-5144}                  & 0.031*                  & \multicolumn{1}{c|}{Education}                                                  & \multicolumn{1}{c|}{-4.515}      & 0.063       \\ \hline
\multicolumn{1}{|c|}{\textbf{\begin{tabular}[c]{@{}c@{}}Neutral\\ sentiment\end{tabular}}}  & \multicolumn{1}{c|}{-5176}                  & 0.03*                   & \multicolumn{1}{c|}{Post length}                                                & \multicolumn{1}{c|}{0.004}       & 0.08        \\ \hline
\multicolumn{1}{|c|}{\textbf{\begin{tabular}[c]{@{}c@{}}Negative\\ sentiment\end{tabular}}} & \multicolumn{1}{c|}{-5138}                  & 0.031*                  & \multicolumn{1}{c|}{\textbf{Readability}}                                       & \multicolumn{1}{c|}{0.391}       & 0.002**     \\ \hline
\multicolumn{1}{|c|}{Post length}                                                           & \multicolumn{1}{c|}{-0.243}                 & 0.058                   & \multicolumn{1}{c|}{\begin{tabular}[c]{@{}c@{}}First-person count\end{tabular}} & \multicolumn{1}{c|}{-0.232}      & 0.093       \\ \hline
\multicolumn{1}{|c|}{First-person count}                                                      & \multicolumn{1}{c|}{-0.243}                 & 0.103                   & \multicolumn{1}{c|}{\textbf{Topic 2}}                                           & \multicolumn{1}{c|}{-87.49}      & 2.64e-11*** \\ \hline
\multicolumn{1}{|c|}{\textbf{Topic 3}}                                                      & \multicolumn{1}{c|}{52.75}                  & 1.77e-06***             & \multicolumn{1}{c|}{Topic 4}                                                    & \multicolumn{1}{c|}{-455.2}      & 0.121       \\ \hline
\multicolumn{1}{|c|}{Topic 8}                                                               & \multicolumn{1}{c|}{21.22}                  & 0.076                   & \multicolumn{1}{c|}{Topic 5}                                                    & \multicolumn{1}{c|}{-26.51}      & 0.098       \\ \hline
\multicolumn{1}{|c|}{\multirow{3}{*}{\textbf{Topic 9}}}                                     & \multicolumn{1}{c|}{\multirow{3}{*}{39.81}} & \multirow{3}{*}{0.016*} & \multicolumn{1}{c|}{\textbf{Topic 8}}                                           & \multicolumn{1}{c|}{-26.84}      & 0.029*      \\ \cline{4-6} 
\multicolumn{1}{|c|}{}                                                                      & \multicolumn{1}{c|}{}                       &                         & \multicolumn{1}{c|}{\textbf{Topic 12}}                                          & \multicolumn{1}{c|}{-97.11}      & 0.0004***   \\ \cline{4-6} 
\multicolumn{1}{|c|}{}                                                                      & \multicolumn{1}{c|}{}                       &                         & \multicolumn{1}{c|}{Topic 14}                                                   & \multicolumn{1}{c|}{-369.8}      & 0.121       \\ \hline
\end{tabular}
}
\begin{tablenotes}
      \small
      \item Signif. codes: ‘***’ 0.001 ‘**’ 0.01 ‘*’ 0.05
\end{tablenotes}
\caption{Regression results for RQ3 (submission scores). \changed{Variables that were not useful in prediction based on stepwise regression were omitted.}}
\label{tab:rq3_regression_submissionscores}
\end{table}

\begin{table}[]
{\small
\begin{tabular}{|ccc|ccc|}
\hline
\multicolumn{3}{|c|}{\textbf{Informational}}                                                                                        & \multicolumn{3}{c|}{\textbf{Emotional}}                                                                                                   \\ \hline
\multicolumn{1}{|c|}{Variable}                                                     & \multicolumn{1}{c|}{Coefficient} & \emph{p}           & \multicolumn{1}{c|}{Variable}                                                     & \multicolumn{1}{c|}{Coefficient} & \emph{p}                  \\ \hline
\multicolumn{1}{|c|}{\textbf{Age}}                                                 & \multicolumn{1}{c|}{55.091}      & 0.032*      & \multicolumn{1}{c|}{Age}                                                          & \multicolumn{1}{c|}{31.01}       & 0.083              \\ \hline
\multicolumn{1}{|c|}{\textbf{Education}}                                           & \multicolumn{1}{c|}{81.871}      & 0.006**     & \multicolumn{1}{c|}{\begin{tabular}[c]{@{}c@{}}Positive\\ sentiment\end{tabular}} & \multicolumn{1}{c|}{23030}       & 0.123              \\ \hline
\multicolumn{1}{|c|}{Marital status}                                               & \multicolumn{1}{c|}{41.874}      & 0.094       & \multicolumn{1}{c|}{\begin{tabular}[c]{@{}c@{}}Neutral\\ sentiment\end{tabular}}  & \multicolumn{1}{c|}{23070}       & 0.122              \\ \hline
\multicolumn{1}{|c|}{\begin{tabular}[c]{@{}c@{}}Negative\\ sentiment\end{tabular}} & \multicolumn{1}{c|}{-430.695}    & 0.086       & \multicolumn{1}{c|}{\begin{tabular}[c]{@{}c@{}}Negative\\ sentiment\end{tabular}} & \multicolumn{1}{c|}{22670}       & 0.129              \\ \hline
\multicolumn{1}{|c|}{\textbf{First-person count}}                                    & \multicolumn{1}{c|}{4.509}       & 2.54e-12*** & \multicolumn{1}{c|}{\textbf{Post length}}                                         & \multicolumn{1}{c|}{0.064}       & \textless 2e-16*** \\ \hline
\multicolumn{1}{|c|}{\textbf{Topic 3}}                                             & \multicolumn{1}{c|}{274.59}      & 0.004**     & \multicolumn{1}{c|}{Topic 2}                                                      & \multicolumn{1}{c|}{227}         & 0.065              \\ \hline
\multicolumn{1}{|c|}{\textbf{Topic 8}}                                             & \multicolumn{1}{c|}{341.721}     & 1.77e-06*** & \multicolumn{1}{c|}{\textbf{Topic 3}}                                             & \multicolumn{1}{c|}{378.2}       & 2.88e-05***        \\ \hline
\multicolumn{1}{|c|}{\textbf{Topic 10}}                                             & \multicolumn{1}{c|}{5542.607}    & 0.031*      & \multicolumn{1}{c|}{\textbf{Topic 8}}                                             & \multicolumn{1}{c|}{310.7}       & 0.006**            \\ \hline
\multicolumn{1}{|c|}{Topic 11}                                                     & \multicolumn{1}{c|}{1474.25}     & 0.144       & \multicolumn{1}{c|}{\textbf{Topic 9}}                                             & \multicolumn{1}{c|}{336.5}       & 0.002**            \\ \hline
\end{tabular}
}
\begin{tablenotes}
      \small
      \item Signif. codes: ‘***’ 0.001 ‘**’ 0.01 ‘*’ 0.05
\end{tablenotes}
\caption{Regression results for RQ3 (comment length). \changed{Variables that were not useful in prediction based on stepwise regression were omitted.}}
\label{tab:rq3_regression_commentlength}
\end{table}

\begin{table}[]
{\small
\begin{tabular}{|ccc|ccc|}
\hline
\multicolumn{3}{|c|}{\textbf{Informational}}                                                                                                                         & \multicolumn{3}{c|}{\textbf{Emotional}}                                                                                                     \\ \hline
\multicolumn{1}{|c|}{Variable}                                                              & \multicolumn{1}{c|}{Coefficient}             & \emph{p}                       & \multicolumn{1}{c|}{Variable}                                                              & \multicolumn{1}{c|}{Coefficient} & \emph{p}           \\ \hline
\multicolumn{1}{|c|}{\textbf{Education}}                                                    & \multicolumn{1}{c|}{0.081}                   & 0.022*                  & \multicolumn{1}{c|}{\textbf{Age}}                                                          & \multicolumn{1}{c|}{-0.064}      & 0.014*      \\ \hline
\multicolumn{1}{|c|}{\textbf{\begin{tabular}[c]{@{}c@{}}Health\\ information\end{tabular}}} & \multicolumn{1}{c|}{0.088}                   & 0.002**                 & \multicolumn{1}{c|}{Education}                                                             & \multicolumn{1}{c|}{0.045}       & 0.107       \\ \hline
\multicolumn{1}{|c|}{\textbf{\begin{tabular}[c]{@{}c@{}}Positive\\ sentiment\end{tabular}}} & \multicolumn{1}{c|}{-2.008}                  & 3.15e-07***             & \multicolumn{1}{c|}{\textbf{Gender}}                                                       & \multicolumn{1}{c|}{0.118}       & 0.001**     \\ \hline
\multicolumn{1}{|c|}{\textbf{\begin{tabular}[c]{@{}c@{}}Neutral\\ sentiment\end{tabular}}}  & \multicolumn{1}{c|}{-1.121}                  & 0.0003***               & \multicolumn{1}{c|}{\textbf{\begin{tabular}[c]{@{}c@{}}Positive\\ sentiment\end{tabular}}} & \multicolumn{1}{c|}{-1.317}      & 6.19e-06*** \\ \hline
\multicolumn{1}{|c|}{\textbf{Topic 1}}                                                      & \multicolumn{1}{c|}{1.979}                   & 0.006**                 & \multicolumn{1}{c|}{\textbf{\begin{tabular}[c]{@{}c@{}}Neutral\\ sentiment\end{tabular}}}  & \multicolumn{1}{c|}{-0.671}      & 0.005**     \\ \hline
\multicolumn{1}{|c|}{\textbf{Topic 3}}                                                      & \multicolumn{1}{c|}{0.387}                   & 0.0007***               & \multicolumn{1}{c|}{\textbf{Readability}}                                                  & \multicolumn{1}{c|}{-0.002}      & 0.035*      \\ \hline
\multicolumn{1}{|c|}{Topic 10}                                                               & \multicolumn{1}{c|}{4.949}                   & 0.113                   & \multicolumn{1}{c|}{\textbf{\begin{tabular}[c]{@{}c@{}}First-person\\ count\end{tabular}}}   & \multicolumn{1}{c|}{0.001}       & 0.013*      \\ \hline
\multicolumn{1}{|c|}{\multirow{8}{*}{\textbf{Topic 7}}}                                     & \multicolumn{1}{c|}{\multirow{8}{*}{-2.822}} & \multirow{8}{*}{0.042*} & \multicolumn{1}{c|}{Topic 1}                                                               & \multicolumn{1}{c|}{1.191}       & 0.137       \\ \cline{4-6} 
\multicolumn{1}{|c|}{}                                                                      & \multicolumn{1}{c|}{}                        &                         & \multicolumn{1}{c|}{\textbf{Topic 2}}                                                      & \multicolumn{1}{c|}{1.187}       & 0.015*      \\ \cline{4-6} 
\multicolumn{1}{|c|}{}                                                                      & \multicolumn{1}{c|}{}                        &                         & \multicolumn{1}{c|}{\textbf{Topic 3}}                                                      & \multicolumn{1}{c|}{1.189}       & 0.011*      \\ \cline{4-6} 
\multicolumn{1}{|c|}{}                                                                      & \multicolumn{1}{c|}{}                        &                         & \multicolumn{1}{c|}{\textbf{Topic 5}}                                                      & \multicolumn{1}{c|}{1.255}       & 0.007**     \\ \cline{4-6} 
\multicolumn{1}{|c|}{}                                                                      & \multicolumn{1}{c|}{}                        &                         & \multicolumn{1}{c|}{\textbf{Topic 8}}                                                      & \multicolumn{1}{c|}{1.057}       & 0.024*      \\ \cline{4-6} 
\multicolumn{1}{|c|}{}                                                                      & \multicolumn{1}{c|}{}                        &                         & \multicolumn{1}{c|}{\textbf{Topic 9}}                                                      & \multicolumn{1}{c|}{1.182}       & 0.014*      \\ \cline{4-6} 
\multicolumn{1}{|c|}{}                                                                      & \multicolumn{1}{c|}{}                        &                         & \multicolumn{1}{c|}{\textbf{Topic 12}}                                                     & \multicolumn{1}{c|}{1.419}       & 0.005**     \\ \cline{4-6} 
\multicolumn{1}{|c|}{}                                                                      & \multicolumn{1}{c|}{}                        &                         & \multicolumn{1}{c|}{Topic 14}                                                              & \multicolumn{1}{c|}{3.587}       & 0.064       \\ \hline
\end{tabular}
}
\begin{tablenotes}
      \small
      \item Signif. codes: ‘***’ 0.001 ‘**’ 0.01 ‘*’ 0.05
\end{tablenotes}
\caption{Regression results for RQ3 (sentiment scores). \changed{Variables that were not useful in prediction based on stepwise regression were omitted.}}
\label{tab:rq3_regression_sentiment}
\end{table}

\vspace{2mm}
\noindent \textbf{Post Attributes and User Engagement} 
For dependent variables, we use all comment features (submission scores/number of comments/number of unique commenters/length of comments/sentiment scores) described in RQ2. This time, however, a total number of comments and unique commenters is aggregated into one variable for simplicity's sake. Given 2 support categories and 4 user engagement features, 8 stepwise regression analyses are performed to filter insignificant predictor variables. 

\begin{itemize}
\setlength\itemsep{1mm}
 \item \changed{\textbf{comment \& commenter count:}}
As shown in Table \ref{tab:rq3_regression_comment}, for informational support, 21 variables are statistically affiliated with the number of comments and commenters. While post authors' disclosure of age and employment information, posts that are longer or discuss seasonal shifts in the COVID-19 outbreak (topic 7) and personal thoughts/opinions (topic 15) are more likely to obtain a larger number of comments, the remaining variables (\textit{race, location, education, marital status disclosure, post sentiments, authors' mentions of themselves, topic 1, topic 2, topic 5, topic 8, topic 11, topic 12, topic 13}) are associated with fewer comments or commenters. Similar to informational support, emotional support seekers tend to engage with others more frequently through comments when they disclose their age information, write longer posts, or talk about topic 15. Also, their voluntary discourse about race, location, education, and marital status do not positively affect the number of comments they obtain. Yet, a majority of dominant topics do not benefit the level of commenters' engagement.

 \item \changed{\textbf{submission scores:}} 
 \changed{According to Table \ref{tab:rq3_regression_submissionscores}, the variables \emph{employment status and location disclosure, post sentiments, post length, the appearance of first-person pronouns, topic 3, topic 8, and topic 10} are selected to fit the informational support model to the highest extent.} Among these factors, sentiment scores, topic 3, and topic 9 are found to be significant. Authors' emotional expressions adversely affect submission sores, but discussions around their acquaintances and work/class activities are linked to higher scores. In contrast, in terms of emotional support, readability is the only parameter that has a positive connection with submission scores. Meanwhile, \changed{topics relevant to anxieties, families, and outdoor social distancing are statically significant but have negative coefficient values.} 

 \item \changed{\textbf{comment length: }}
A subset of 9 variables \emph{age, education, marital status disclosure, negative sentiments, first-person pronouns, topic 3, topic 8, topic 10, and topic 11} was chosen for informational support \changed{(Table \ref{tab:rq3_regression_commentlength})}. In particular, we find that there exist positive relations between the average length of comments and \emph{age or education disclosure, first-person pronouns, topic 3, topic 8, and topic 10} variables. 
Topic 3 and topic 8 are about families and friends whereas topic 10 is more relevant to wearing masks and social distancing outside. In terms of emotional support, \emph{age information disclosure, three sentiment (positive/negative/neutral) scores, post length, and three topics (topic 3, 8, and 9)} parameters were selected. Similar to informational support seeking posts, we discover substantial connections between comment length and topics 3 and 8 in postings requesting emotional aid. Moreover, as the length of postings grows, so does the length of comments.

 \item \changed{\textbf{sentiment scores: }} 
\newchanged{As shown in Table\ref{tab:rq3_regression_sentiment}}, a combination of 8 parameters \emph{education or health-related disclosure, positive or neutral sentiments, topic 1, topic 3, topic 7, and topic 10} can predict the sentiment levels of comments for informational support-seeking posts most adequately. All variables except for topic 10 are proven to be statistically significant. \changed{Specifically, categories including education/health information disclosure and discussion about cleaning (topic 1) and family members (topic 3) have positive relationships with comment sentiments, posters' emotional expressions and mentions of the coronavirus surge (topic 7) are affiliated with negative emotions expressed within comments.} For emotional support seeking posts, \changed{14 variables were chosen (see Table \ref{tab:rq3_regression_sentiment}), and a majority of them, excluding the education level information sharing, topics 1 and 14, were statistically significant.} 



\end{itemize}

\section{Discussion}
\newchanged{The current study investigates 11 personal information types that users publicly disclose in a sample of conversations on the r/COVID19\_support subreddit. Our goal is to gain a comprehensive understanding of how self-disclosure, support-seeking, and user engagement interplay within a social support context. Following, we discuss our findings in the context of existing research, as well as their practical and ethical implications. We further describe the limitations of our work.}

\vspace{2mm} 
\subsection{Self-Disclosure and Support-Seeking \newchanged{during COVID-19}} 
Consistent with \citet{avizohar2022facebook}'s findings of medical support forums with an increased rate of online self-disclosure, nearly two-thirds of postings contained the posters' private data. Particularly, users' voluntary information sharing about health information (23.59\%) was dominant in comparison to other categories. It is not surprising but congruent with a recent study \cite{saha2020social}. Users also frequently shared personal data for their employment status (19.8\%), marital status (16.63\%), and location (16.09\%). We assume these phenomena are explainable by an increase in discussions around risk factors linked to medical conditions, unemployment or work-related trauma, and varying degree of the severity of the virus in different states. 

Of 11 categories of personal information, differences in online self-disclosure amongst requested support types for age, education, and location categories were statistically significant. More precisely, we notice an increased frequency of personal information sharing practices when users sought both informational and emotional aid, whereas posts that are solely requesting informational support tend to have the lowest instance of observed self-disclosure. This finding indicates that those who seek a greater amount of support divulge their personal information, particularly regarding age, education, and location, more often. Although we cannot speak the base level disclosure for the identified categories before the Coronavirus outbreak, it is well established that indiscriminate online disclosure can pose social stigma and make users more susceptible to cyber attacks \cite{bansal2007impact,patil2012reasons,friedland2010cybercasing}. In essence, our observations raise privacy concerns in the context of, in many cases, already vulnerable individuals seeking support during a public health crisis. 

\vspace{2mm} 
\changed{\subsection{Self-Disclosure and the Receipt of Social Support and User Engagement \newchanged{during COVID-19}}}
Unlike previous works \citep{wingate2020influence, andalibi2016understanding} that analyzed a correlation between public discourse and the amount of support received, we examine how users' data sharing may influence the reciprocity of \emph{solicited} support. Our results reveal that the intensified degree of self-disclosure in the post, as measured by the number of distinct categories of personal information shared, was strongly associated with acquiring the preferred type of informational support, but not emotional support. The observed disparity might be due to respondents' perceptions that people are looking for a specific form of social support. \citet{yang2019channel}, for example, found that sharing unpleasant thoughts and feelings is significantly linked to the notion that people are seeking emotional support, which is subsequently tied to their obtaining emotional support from others. Similarly, users' objective self-disclosure may have led respondents to conjecture that they are pursuing informational assistance. Additional qualitative investigation into why they have provided such support would clarify the question. \changed{In line with \citet{andalibi2018social}'s findings, these results imply that support-related subreddits successfully facilitate informational support exchange. At the same time, this casts doubt on Reddit's capacity to maintain OHC (especially for the COVID-19 virus) because the bulk of posts requested emotional support but failed to get it.}

Further, we test the effect of online self-disclosure on five user engagement measures for two support types independently and confirm that informational and emotional help seeking postings have slightly different associations with them. For both support types, as suggested in \citet{pan2018you}, the increase in self-disclosure rates was strongly affiliated with longer comments and heightened sentiment levels on average. Yet, while self-disclosing posts that pursue informational help were more likely to obtain higher submission scores, those seeking emotional support did not. Three factors (submission scores, the number of comments, and commenters) were negatively associated with personal information sharing with respect to emotional support solicitation. Taking into account the adverse impact of reduced user involvement levels on posters' self-esteem \cite{thomaes2010like}, these findings suggest that not all support-seeking individuals may not profit greatly from their openness, unlike their expectations. Overall, the observed distinctions are meaningful because they allow a research community to reconsider the significance of self-disclosure in obtaining \changed{societal benefits}, particularly for emotional support. 

\vspace{5mm} 
\changed{\subsection{Linguistic Markers of Self-Disclosure and the Receipt of Social Support and User Engagement \newchanged{during COVID-19}}}
Results of RQ3 showed that negative sentiments, the post readability, and the inclusion of first-person pronouns had strong connections with emotional support elicitation, whereas they did not with informational support seeking posts. This partially strengthens prior literature \cite{chen2020linguistic} where they demonstrated a positive correlation between readability and the receipt of social support. In terms of topical variations, both support types successfully obtained the requested support to a greater extent when their content talked about individuals' family members or test results. At the same time, discourse around emotional struggles and their online educational/professional settings tended to successfully receive the desired support for emotional support, but not for emotional support. 


In terms of correlations between post-level attributes and user engagement features, we uncover several noticeable patterns. First, intensified emotions or feelings within posts were prone to receive a lower amount of respondents' involvement (in terms of the number of comments/commenters, submission scores, and sentiment scores) when the posters sought informational assistance. However, the emotional appeal did affect the comments' sentiment levels. These results contradict prior observations of a positive impact of sentiment disclosure on user engagement \cite{bil2022all, molina2020content}.
Second, while the length of emotional support seeking postings and their readability grades are statistically significant factors in several user involvement features, post length was determined to be the only element that affects user engagement, particularly the number of comments and commenters, for informational support.
Third, among different personal information categories, disclosing education level positively influenced several user engagement features (the number of comments/commenters, the length of comments, and their sentiments) for individuals seeking informational support. For emotional support seekers, there was no consistent trend across user involvement features for any disclosure variable.
Lastly, topical variations were correlated with post viewers' engagement for both emotional and informational support seeking content. Discussing anxieties and fear towards the virus or household situations had a favorable impact on the length and sentiment of received comments for informational support seekers. Yet, it was adversely associated with the number of comments, commenters, and post submission scores. In contrast, those seeking informational support received higher submission scores, longer comments, and heightened sentimental reactions when discussing friends or family members.

In essence, our findings suggest that users are more inclined to publicly engage with posts based on specific linguistic attributes. This encourages users to deliberately invest their time and resources, depending on the kind of desired help, to sustain strong linkages with others.

\vspace{5mm} 
\newchanged{\subsection{Implications}
\subsubsection{Practical Implications}
Our work contributes to understanding how COVID-19 support community members acquire social support and user participation through self-disclosure practices. It highlights the trade-offs between perceived rewards (enhancing the utility or usefulness of counseling) and privacy risks (losing control over their data) resulting from public discourse. Although there has been growing evidence regarding the benefits of self-disclosure on the reciprocity of support, our findings call attention to the disproportionate privacy vulnerability of emotional help seekers.

We suggest that OHC designers and moderators investigate innovative ways to assist all support-seeking individuals in exchanging social support with enhanced privacy awareness, drawing from existing research (e.g., \citet{glaeser2005paternalism, beaulier2007behavioral, lampinen2011we}) on their benefits in decision-making. Describing potential privacy harms that may occur when particular pieces of personal data are revealed has been reported to have a positive impact on self-disclosure regulation \cite{marmion2017cognitive, diaz2020preventative}. Similarly, operators of \textbf{r/COVID19\_support} subreddit can provide users comprehensive and personalized instructions on what not to share based on our analyses of support seekers' disclosure patterns. The proposed methodology can also be used to automate this process by training an artificial intelligence (AI) model to analyze post content and estimate privacy leakage and social support acquisition in real time. 

Gaining an understanding of the attributes that foster adequate social support and user engagement is crucial for improving support seekers' psychological coping abilities and physical well-being \citep{macgeorge2011supportive}. In line with prior literature \cite{chen2020linguistic, rains2016language}, we find that linguistic signals significantly affect response acquisition from audiences. Our detailed analyses on RQ3 can help support seekers target specific reaction categories based on their content. In a similar vein, designers and moderators can use the identified patterns to facilitate a constructive social support exchange. 

Finally, our work reflects on how social support is exchanged through the act of online self-disclosure during the pandemic. It affords opportunities to predict the dynamics of public discourse and social support and user engagement provision in future global health crises, as well as motivating audiences to engage effectively with vulnerable self-disclosing individuals.

\subsubsection{Ethical Implications}
For this research, we created new datasets of 2,399 posts and 29,851 associated comments collected from the \textbf{r/COVID19\_support} subreddit where each content is tagged with specific personal information categories and social support categories. Data collection processes were performed by the first author, and all crawled data was stored in the author’s computer with password protection. For annotation and analyses, the first author then uploaded it to the password-protected OneDrive\footnote{OneDrive is a university-contracted storage provider in which two-factor authentication is part of the security} folder, and it was only accessible by authors of this work.

Although key characteristics of Reddit are publicity and anonymity \citep{Gutman2018}, we chose not to share our annotated data due to ethical concerns. Specifically, they explicitly expose what types of personal and private information are included in the content and can be maliciously used by cyber attackers. For this reason, our labels for self-disclosure will not be shared.

The proposed model for support-giving comment classification is trained with data that is downloaded from two subreddits (`r/CasualConversation' and `r/OffMyChest'). Even though our model achieves a better performance than the state-of-the-art method \cite{dadu2020bert}, models trained with data acquired from one setting may suffer from poor generalization issues when implemented in other practice settings \cite{safdar2020ethical}. Given our dataset is crawled from a different subreddit (`r/COVID19\_support'), we reexamined its performance with manual inspection and confirmed its generalizability (see Section \ref{sec:automated}). Yet, it is important to note that we did not test its robustness on data from other social media platforms such as Facebook or Twitter. We highly recommend that those who plan on utilizing our approach reassess their local dataset.
}

\subsection{Limitations}
We discuss \changed{four} limitations of this study. First, our findings target users' self-disclosing behaviors within one particular health support subreddit during the COVID-19 pandemic, and thus may not generalize to other types of online health communities. We note that users may demonstrate different patterns of self-disclosure when they interact with other groups with different purposes or on different social media platforms. Future research can revisit the proposed research inquiries on more diverse support-related forums on different platforms (e.g., Facebook, Twitter). Second, our analysis of the third research question solely relies on the content itself. As part of the future work, it would be interesting to consider other features such as multimedia content (e.g., images and videos) or context (e.g., posted time and location), and the sociodemographic attributes (e.g., age and gender) of a post creator. Third, other types of social support (e.g., instrumental support, esteem support, network support) were excluded from this research. This was done so that we could use publicly available data for establishing gold standard labels and to avoid additional subjectivity. \changed{Finally, we did not examine other deep learning model architectures such as CNN or LSTM for our support classifier. Instead, we primarily focused on transformer-based models like BERT due to their advanced contextual understanding of texts. Future work should explore whether more recently proposed language models, including GPT-2, GPT-3, and BLOOM, can further increase the performance.}

\section{Conclusion}

In our study, we analyze users’ voluntary disclosures, particularly in the context of support-seeking activities during the Coronavirus epidemic. Throughout the investigation of the proposed research questions, we identified the valence of health and occupation information when posters interact with other members of the COVID-19 support group. At the same time, their information-sharing behaviors tended to shift based on what types of support are requested in the post: users were prone to reveal their age, education, and location information more frequently when seeking both informational and emotional support, as opposed to pursuing either one. Nonetheless, it was shown that the benefits of self-disclosure were not evenly distributed. While revealing personal information increased the chances of individuals obtaining requested informational help, the same could not be guaranteed for emotional support. Likewise, it did not positively impact other users' involvement for emotional support seekers whereas informational support seekers obtained longer comments and higher submission or sentiment scores in return for their disclosures. Lastly, we report differing impacts of linguistic factors in postings on support receipt and user involvement by requested support type. The current research not only enriches the understanding of the role of online self-disclosure in social \changed{support and user engagement} acquisition during the COVID-19 pandemic but also sheds light on practical strategies to optimize social support exchange and user engagement by leveraging post-level attributes.


\bibliographystyle{ACM-Reference-Format}
\bibliography{ACMTSC-Jooyoung}

\appendix

\end{document}